\documentclass[journal,12pt,onecolumn]{IEEEtran}

\usepackage{setspace}
\ifCLASSOPTIONonecolumn \doublespace \fi
\usepackage{graphics}
\usepackage{rotating}
\usepackage{amsfonts}
\usepackage{amsmath}
\usepackage{subfigure}
\usepackage{multirow}
\allowdisplaybreaks[1]
\newtheorem{theorem}{Theorem}
\renewcommand{\IEEEQED}{\IEEEQEDopen}

\graphicspath{{Figures/}} \DeclareGraphicsExtensions{.eps,.ps}

\begin{document}

\title{Constellation Precoded Multiple Beamforming}

\ifCLASSOPTIONconference
\author{\IEEEauthorblockN{Hong Ju Park, Boyu Li, and Ender Ayanoglu}\\
\IEEEauthorblockA{Center for Pervasive Communications and Computing\\
Department of Electrical Engineering and Computer Science\\
University of California, Irvine\\
Email: hjpark@uci.edu, boyul@uci.edu, ayanoglu@uci.edu}} \else
\author{Hong~Ju~Park,~\IEEEmembership{Student~Member,~IEEE,}
Boyu~Li,~and~Ender~Ayanoglu,~\IEEEmembership{Fellow,~IEEE}%
\thanks{H. J. Park, B. Li, and E. Ayanoglu are with the Center for Pervasive
Communications and Computing, Department of Electrical Engineering
and Computer Science, Henry Samueli School of Engineering,
University of California, Irvine, CA 92697-3975 USA (e-mail:
hjpark@uci.edu; boyul@uci.edu; ayanoglu@uci.edu).}} \fi

\maketitle

\ifCLASSOPTIONonecolumn
 \setlength\arraycolsep{4pt}
\else
 \setlength\arraycolsep{2pt}
 \def\sizefig{0.95}
\fi

\begin{abstract}
\boldmath Beamforming techniques that employ Singular Value
Decomposition (SVD) are commonly used in Multi-Input Multi-Output
(MIMO) wireless communication systems. In the absence of channel
coding, when a single symbol is transmitted, these systems achieve
the full diversity order provided by the channel; whereas when
multiple symbols are simultaneously transmitted, this property is
lost. When channel coding is employed, full diversity order can be
achieved. For example, when Bit-Interleaved Coded Modulation (BICM)
is combined with this technique, full diversity order of $NM$ in an
$M \times N$ MIMO channel transmitting $S$ parallel streams is
possible, provided a condition on $S$ and the BICM convolutional
code rate is satisfied.  In this paper, we present constellation
precoded multiple beamforming which can achieve the full diversity
order both with BICM-coded and uncoded SVD systems. We provide an
analytical proof of this property. To reduce the computational
complexity of Maximum Likelihood (ML) decoding in this system, we
employ Sphere Decoding (SD). We report an SD technique that reduces
the computational complexity beyond commonly used approaches to SD.
This technique achieves several orders of magnitude reduction in
computational complexity not only with respect to conventional ML
decoding but also, with respect to conventional SD.
\end{abstract}


\section{Introduction} \label{sec:introduction}

When the perfect channel state information is available at the
transmitter, beamforming is employed to achieve spatial multiplexing
and thereby increase the data rate, or to enhance the performance of
a Multiple-Input Multiple-Output (MIMO) system
\cite{jafarkhaniBook}. The beamforming vectors are designed in
\cite{SampathJCOM01}, \cite{palomarTSP03} for various design
criteria, and can be obtained by the Singular Value Decomposition
(SVD), leading to a channel-diagonalizing structure optimum in
minimizing the average Bit Error Rate (BER) \cite{palomarTSP03}.
Uncoded Single Beamforming (SB), which carries only one symbol at a
time, was shown to achieve the full diversity order of $NM$ where
$N$ and $M$ are the number of transmit and receive antennas,
respectively \cite{sengulTC06AnalSingleMultpleBeam},
\cite{OrdonezTSP07}. However, the diversity order of uncoded
multiple beamforming, which increases the throughput by sending
multiple symbols at a time, is $(N -S + 1)(M - S + 1)$ where the
symbols are transmitted on the subchannels with the largest $S$
singular values, losing the full diversity order over flat fading
channel \cite{sengulTC06AnalSingleMultpleBeam}, \cite{OrdonezTSP07}.

It is known that an SVD subchannel with larger singular value
provides larger diversity gain \cite{OrdonezTSP07}. Under the
simultaneous parallel transmission of the symbols on the
diagonalized subchannels, the performance at high Signal-to-Noise
Ratio (SNR) is dominated by the subchannel with the smallest
singular value. To overcome the degradation of the diversity order
of multiple beamforming, Bit-Interleaved Coded Multiple Beamforming
(BICMB) was proposed \cite{akayTC06BICMB},
\cite{akayTC06BICMB_arxiv}. This scheme interleaves the codewords
through the multiple subchannels with different diversity order,
resulting in better diversity order. BICMB can achieve the full
diversity order offered by the channel as long as the code rate
$R_c$ and the number of employed subchannels $S$ satisfy the
condition $R_c S \leq 1$ \cite{ParkICC09}.

In this paper, we present a multiple beamforming technique that
achieves the full diversity order in both of the coded and the
uncoded systems. This technique employs the constellation precoding
scheme \cite{GamalJIT03}, \cite{XinJWCOM03}, \cite{LiuJCOM03},
\cite{ZhangJCOM07}, \cite{GressetGlobecom09}, which is used for
space-time or space-frequency block codes to increase the system
data rate without losing the full diversity order. We show via
Pairwise Error Probability (PEP) analysis that Fully Precoded
Multiple Beamforming (FPMB) with Maximum Likelihood (ML) detection
achieves the full diversity order even in the absence of any channel
coding. We also present the diversity analysis of Bit-Interleaved
Coded Multiple Beamforming with Constellation Precoding (BICMB-CP),
which adds the constellation precoding stage to BICMB. We show that
the addition of the constellation precoder to BICMB, whose code rate
$R_c$ is larger than $1/S$, provides the full diversity when the
subchannels for the precoded symbols are properly chosen. Simulation
results are shown to prove the analysis.

Multiple beamforming without constellation precoding separates the
MIMO channel into independent parallel subchannels, enabling
symbol-by-symbol detection on each subchannel. Since the precoder at
the transmitter no longer allows the parallel independent detection
of the symbols on each subchannel, the complexity of the ML
detection for precoded symbols, which provides optimal performance,
increases exponentially with the number of possible constellation
points of the modulation scheme and the dimension of the
constellation precoder. The complexity increase makes the receiver
with the ML detection unsuitable for practical purposes
\cite{ZimmermannWPMC04}. On the other hand, Sphere Decoding (SD) was
proposed as an alternative for ML detection that provides optimal
performance with reduced computational complexity
\cite{JaldenJSP05}.

Several complexity reduction techniques for SD have been proposed.
In \cite{HanGLOBECOM05} and \cite{ChengISCC07}, attention is drawn
to the initial radius selection strategy, since an inappropriate
initial radius can result in either a large number of lattice points
to be searched, or a number of restarted searches with increased
initial radius. In \cite{HassibiJSP05} and \cite{ZhaoJCOM05}, the
complexity is reduced by making a proper choice to update the sphere
radius. Other methods, such as the $K$-best lattice decoder
\cite{WongISCAS02}, \cite{HuynhISWCS08}, and a combination of SD and
$K$-best decoder \cite{TangICC04}, can significantly reduce the
complexity of low SNR at the cost of BER performance.

In this paper, we propose an SD algorithm which efficiently improves
the complexity of constellation precoded multiple beamforming over
flat fading channel by reducing the average number of
multiplications required to obtain the optimal solution. This
complexity reduction is accomplished by precalculating the
multiplications at the beginning of decoding, and recycling them
later for the repetitive calculations. The reduction is achieved
further by the help of the lattice representation of our previous
work presented in \cite{AzzamGLOBECOM07}, which introduces
orthogonality between the real and imaginary parts of every detected
symbol. Based on Zero-Forcing Decision Feedback Equalization
(ZF-DFE), the proposed SD algorithm includes a method to determine
the initial radius, reducing the average number of real
multiplications needed to acquire one precoded bit metric for
BICMB-CP. With simulation results, we show that conventional SD
reduces the complexity substantially compared with the exhaustive
search, and the complexity can be further reduced effectively by our
proposed SD. The complexity reduction becomes larger as the
constellation precoder dimension and the constellation size become
larger.

The rest of this paper is organized as follows. The description of
uncoded and coded multiple beamforming combined with constellation
precoding is given in Section \ref{sec:system_model}. Sections
\ref{sec:div_anal_UMB-CP} and \ref{sec:div_anal_BICMB-CP} present
the diversity analysis of the MIMO schemes through the calculation
of the upper bound to PEP. The computational complexity reduction
sphere detection algorithm is discussed in Section \ref{sec:sphere}.
Simulation results supporting the analysis are shown in Section
\ref{sec:results}. Finally, we end the paper with our conclusion in
Section \ref{sec:conclusion}.

\textbf{Notation:} Bold lower (upper) case letters denote vectors
(matrices). $\textrm{diag}[\mathbf{B}_1, \cdots, \mathbf{B}_P]$
stands for a block diagonal matrix with matrices $\mathbf{B}_1,
\cdots, \mathbf{B}_P$, and $\textrm{diag}[b_1, \cdots, b_P]$ is a
diagonal matrix with diagonal entries $b_1, \cdots, b_P$.
$\Re(\cdot)$ and $\Im(\cdot)$ denote the real and imaginary part of
a complex number, respectively. The superscripts $(\cdot)^H$,
$(\cdot)^T$, $(\cdot)^*$, $\bar{(\cdot)}$ stand for conjugate
transpose, transpose, complex conjugate, binary complement,
respectively, and $\forall$ denotes for-all. $\lceil \cdot \rceil$
is the ceiling function that maps a real number to the next largest
integer. $\mathbb{R}^+$ and $\mathbb{C}$ stand for the set of
positive real numbers and the complex numbers, respectively.
$d_{min}$ is the minimum Euclidean distance between two points in a
constellation.

\section{System Model} \label{sec:system_model}

\subsection{Uncoded Multiple Beamforming with Constellation Precoding}
\label{sec:system_model_uncoded}
Uncoded Multiple Beamforming with
Constellation Precoding (UMB-CP) transforms modulated symbols to
precoded symbols via a precoding matrix. The $S \times 1$ symbol
vector $\mathbf{x}$, where $S \leq \min(N, M)$, is precoded by a
square matrix $\mathbf{\Theta}$. We assume that the elements of
$\mathbf{x}$ belong to a signal set $\chi \subset \mathbb{C}$ of
size $|\chi| = 2^m$, such as $2^m$-QAM, where $m$ is the number of
input bits to the Gray encoder. The precoder is expressed as
\begin{align}
\mathbf{\Theta} = \left[ \begin{array}{cc}
\mathbf{\tilde{\Theta}} & \mathbf{0} \\
\mathbf{0} & \mathbf{I}_{S-P}
\end{array} \right]
\label{eq:precoder_def}
\end{align}
where $\mathbf{\tilde{\Theta}}$ is a $P \times P$ constellation
precoding matrix that precodes the first $P$ modulated symbols of
the vector $\mathbf{x}$. When all of the $S$ modulated symbols are
precoded ($P = S$), we call the resulting system Fully Precoded
Multiple Beamforming (FPMB), otherwise, we call it Partially
Precoded Multiple Beamforming (PPMB). The permutation matrix
$\mathbf{T}$ reorders the precoded $P$ symbols and non-precoded
$S-P$ symbols to be transmitted on the predefined subchannels
created by the SVD of the MIMO channels. Let us define
$\boldsymbol{\eta} = \left[ \eta_1 \, \cdots \, \eta_P \right]$ as a
vector whose element $\eta_p$ is the index of the subchannel on
which the precoded symbols are transmitted, and ordered increasingly
such that $\eta_p < \eta_q$ for $p < q$. In the same way,
$\boldsymbol{\omega} = \left[ \omega_1 \, \cdots \, \omega_{(S-P)}
\right]$ is defined as an increasingly ordered vector whose elements
are the indices of the subchannels which carry the non-precoded
symbols.

The MIMO channel $\mathbf{H} \in \mathbb{C}^{M \times N}$ is assumed
to be quasi-static, Rayleigh, and flat fading, and perfectly known
to both the transmitter and the receiver. The beamforming matrices
are determined by the SVD of the MIMO channel, i.e., $\mathbf{H} =
\mathbf{U \Lambda V^H}$ where $\mathbf{U}$ and $\mathbf{V}$ are
unitary matrices, and $\mathbf{\Lambda}$ is a diagonal matrix whose
$s^{th}$ diagonal element, $\lambda_s \in \mathbb{R}^+$, is a
singular value of $\mathbf{H}$ in decreasing order. When $S$ symbols
are transmitted at the same time, then the first $S$ vectors of
$\mathbf{U}$ and $\mathbf{V}$ are chosen to be used as beamforming
matrices at the receiver and the transmitter, respectively. In
\figurename{ \ref{fig:system_model_uncoded}} which displays the
structure of UMB-CP, $\mathbf{\tilde{U}}$ and $\mathbf{\tilde{V}}$
denote the beamforming matrices picked from $\mathbf{U}$ and
$\mathbf{V}$.

\ifCLASSOPTIONonecolumn
\begin{figure}[!m]
\centering{ \subfigure[Uncoded Multiple Beamforming with
Constellation
Precoding.]{\includegraphics[width=0.6\linewidth]{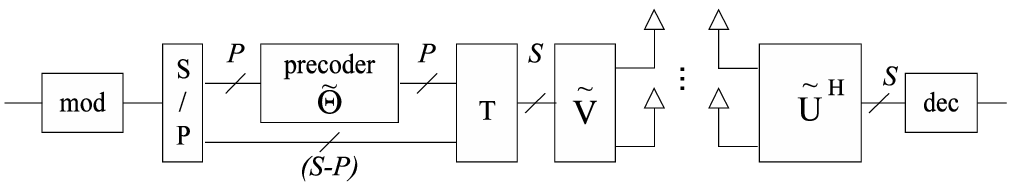}
\label{fig:system_model_uncoded}} \hfil \subfigure[Bit-Interleaved
Coded Multiple Beamforming with Constellation
Precoding.]{\includegraphics[width=0.6\linewidth]{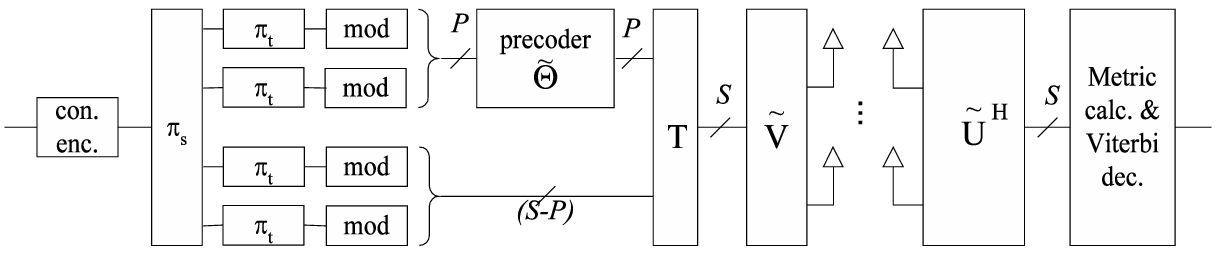}
\label{fig:system_model_coded}}} \caption{Structure of Constellation
Precoded Multiple Beamforming.} \label{fig:structure_CPMB}
\end{figure}
\else
\begin{figure}[!t]
\centering{ \subfigure[Uncoded Multiple Beamforming with
Constellation
Precoding.]{\includegraphics[width=\sizefig\linewidth]{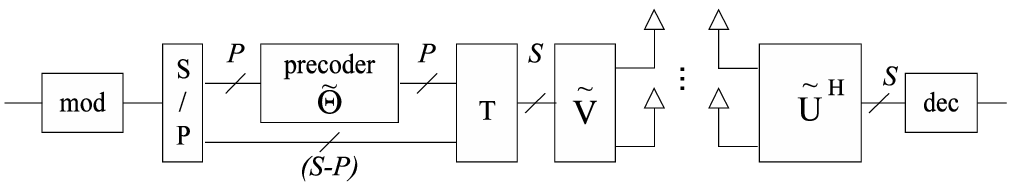}
\label{fig:system_model_uncoded}} \hfil \subfigure[Bit-Interleaved
Coded Multiple Beamforming with Constellation
Precoding.]{\includegraphics[width=\sizefig\linewidth]{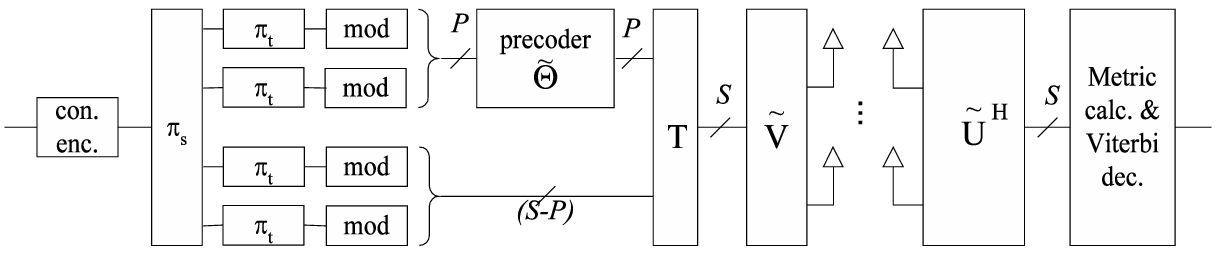}
\label{fig:system_model_coded}}} \caption{Structure of Constellation
Precoded Multiple Beamforming.} \label{fig:structure_CPMB}
\end{figure}
\fi

The serial-to-parallel converter organizes the symbol vector
$\mathbf{x}$ as $\mathbf{x} = [\mathbf{x}_{\boldsymbol{\eta}}^T \,
\vdots \, \mathbf{x}_{\boldsymbol{\omega}}^T]^T = [x_{\eta_1} \,
\cdots \, x_{\eta_P} \vdots$ $\, x_{\omega_1} \, \cdots \,
x_{\omega_{(S-P)}}]^T$, where $\mathbf{x}_{\boldsymbol{\eta}}$ and
$\mathbf{x}_{\boldsymbol{\omega}}$ consist of the modulated entries
to be transmitted on the subchannels specified in
$\boldsymbol{\eta}$ and $\boldsymbol{\omega}$, respectively. The $S
\times 1$ detected symbol vector $\mathbf{y}  = [ \mathbf{y}_p^T \,
\vdots \, \mathbf{y}_n^T]^T = [y_{1} \, \cdots \, y_{P} \, \vdots$
$\, y_{P+1} \, \cdots \, y_{S}]^T$ at the receiver is written as
\begin{align}
\mathbf{y} = \mathbf{\Gamma} \mathbf{\Theta x} + \mathbf{n}
\label{eq:detected_PMB}
\end{align}
where $\boldsymbol{\Gamma}$ is a block diagonal matrix,
$\boldsymbol{\Gamma} = \textrm{diag}[\boldsymbol{\Gamma}_p, \,
\boldsymbol{\Gamma}_n]$, with diagonal matrices defined as
$\boldsymbol{\Gamma}_p = \textrm{diag}[\lambda_{\eta_1}, $ $\,
\cdots, \, \lambda_{\eta_P}]$, $\boldsymbol{\Gamma}_n =
\textrm{diag}[\lambda_{\omega_1}, \, \cdots, \,
\lambda_{\omega_{(S-P)}}]$, and $\mathbf{n} = [ \mathbf{n}_p^T \,
\vdots \, \mathbf{n}_n^T]^T$ is an additive white Gaussian noise
vector with zero mean and variance $N_0 = N / SNR$. The matrix
$\mathbf{H}$ is complex Gaussian with zero mean and unit variance,
and to make the received signal-to-noise ratio $SNR$, the total
transmitted power is scaled as $N$. The input-output relation in
(\ref{eq:detected_PMB}) is decomposed into two equations as
\begin{equation}
\begin{split}
\mathbf{y}_p = \boldsymbol{\Gamma}_p \boldsymbol{\tilde{\Theta
}} \mathbf{x}_{\boldsymbol{\eta}} + \mathbf{n}_p \\
\mathbf{y}_n = \boldsymbol{\Gamma}_n
\mathbf{x}_{\boldsymbol{\omega}} + \mathbf{n}_n.
\label{eq:deteced_symbol_decomposed_uncoded}
\end{split}
\end{equation}
The ML decoding of the detected symbol $\mathbf{\hat{x}} =
[\mathbf{\hat{x}}_{\boldsymbol{\eta}}^T \, \vdots \,
\mathbf{\hat{x}}_{\boldsymbol{\omega}}^T]^T = [\hat{x}_{\eta_1} \,
\cdots \, \hat{x}_{\eta_P} \, \vdots \, \hat{x}_{\omega_{1}} \,
\cdots \, \hat{x}_{\omega_{(S-P)}}]^T$ is given by
\begin{align}
\mathbf{\hat{x}} = \arg \min_{\mathbf{x} \in \chi^S} \left\|
\mathbf{y} - \mathbf{\Gamma} \mathbf{\Theta x} \right\|^2
\label{eq:MLdecoding_PMB}
\end{align}
where $\chi^S$ represents the $S$-dimensional product space based on
$\chi$. For PPMB, the symbol can be detected in a parallel fashion
as
\begin{align}
\mathbf{\hat{x}}_{\boldsymbol{\eta}} = \arg \min_{\mathbf{x} \in
\chi^P} \left\| \mathbf{y}_p - \mathbf{\Gamma}_p
\mathbf{\tilde{\Theta} x} \right\|^2
\label{eq:MLdecoding_PPMB_precoded}
\end{align}
for the precoded symbol, and
\begin{align}
\hat{x}_l = \arg \min_{x \in \chi} | y_l - \lambda_{\tilde{l}} x |^2
\label{eq:MLdecoding_PPMB_non_precoded}
\end{align}
for the non-precoded symbol where $\tilde{l}$ is the corresponding
index transformed by $\mathbf{T}$.

\subsection{Bit-Interleaved Coded Multiple Beamforming with Constellation Precoding}
\label{sec:system_model_coded}
\figurename{
\ref{fig:system_model_coded}} represents the structure of
Bit-Interleaved Coded Multiple Beamforming with Constellation
Precoding (BICMB-CP). First, the convolutional encoder with code
rate $R_c = k_c / n_c$, possibly combined with a perforation matrix
for a high rate punctured code, generates the codeword $\mathbf{c}$
from the information bits. Then, the spatial interleaver $\pi_s$
distributes the coded bits into $S$ streams, each of which is
interleaved by an independent bit-wise interleaver $\pi_t$. The
interleaved bits are mapped by Gray encoding onto the symbol
sequence $\mathbf{X} = [\mathbf{x}_1 \, \cdots \, \mathbf{x}_K]$,
where $\mathbf{x}_k$ is an $S \times 1$ symbol vector at the
$k^{th}$ time instant. Each entry of $\mathbf{x}_k$ belongs to a
signal set $\chi$.

The symbol vector $\mathbf{x}_k$ is multiplied by the $S \times S$
precoder $\boldsymbol{\Theta}$ in (\ref{eq:precoder_def}). When all
of the $S$ modulated entries are precoded ($P = S$), we call the
resulting system Bit-Interleaved Coded Multiple Beamforming with
Full Precoding (BICMB-FP), otherwise, we call it Bit-Interleaved
Coded Multiple Beamforming with Partial Precoding (BICMB-PP). The
precoded symbol vector is transmitted on the MIMO channel described
in Section \ref{sec:system_model_uncoded}.

As in UMB-CP, the spatial interleaver arranges the symbol vector
$\mathbf{x}_k$ as $\mathbf{x}_k = [\mathbf{x}_{k,
\boldsymbol{\eta}}^T \, \vdots \, \mathbf{x}_{k,
\boldsymbol{\omega}}^T]^T = [x_{k,\eta_1}$ $\cdots \, x_{k,\eta_P}
\, \vdots \, x_{k,\omega_1} \, \cdots \, x_{k, \omega_{(S-P)}}]^T$.
The $S \times 1$ detected symbol vector $\mathbf{r}_k = [
(\mathbf{r}_{k}^p)^T \, \vdots \, (\mathbf{r}_{k}^n)^T]^T = [r_{k,1}
\, \cdots$ $r_{k, P} \, \vdots \, r_{k, P+1} \, \cdots \,
r_{k,S}]^T$ at the $k^{th}$ time instant is
\begin{align}
\mathbf{r}_k = \boldsymbol{\Gamma} \mathbf{\Theta x}_k +
\mathbf{n}_k \label{eq:detected_symbol}
\end{align}
where $\mathbf{n}_k = [ (\mathbf{n}_{k}^p)^T \, \vdots \,
(\mathbf{n}_{k}^n)^T]^T$ is an additive white Gaussian noise vector.

The location of the coded bit $c_{k'}$ within the symbol sequence
$\mathbf{X}$ is known as $k' \rightarrow (k, l, i)$, where $k$, $l$,
and $i$ are the time instant in $\mathbf{X}$, the symbol position in
$\mathbf{x}_k$, and the bit position on the label $x_{k,l}$,
respectively. Let $\chi_{b}^{i}$ denote a subset of $\chi$ whose
labels have $b \in \{0, 1\}$ in the $i^{th}$ bit position. By using
the location information and the input-output relation in
(\ref{eq:detected_symbol}), the receiver calculates the maximum
likelihood bit metrics for the coded bit $c_{k'}$ as
\begin{align}
\gamma^{l,i}(\mathbf{r}_{k}, c_{k'}) = \min_{\mathbf{x} \in
\xi_{c_{k'}}^{l,i}} \| \mathbf{r}_{k} - \boldsymbol{\Gamma}
\boldsymbol{\Theta} \mathbf{x} \|^2 \label{eq:ML_bit_metrics}
\end{align}
where $\xi_{c_{k'}}^{l,i}$ is a subset of $\chi^S$, defined as
\begin{align*}
\xi_{b}^{l,i} = \{ \mathbf{x} = [x_1 \, \cdots \, x_S ]^T :
x_{s|s=l} \in \chi_{b}^{i}, \textrm{ and } x_{s|s \neq l} \in \chi
\}.
\end{align*}
In particular, based on the decomposition of
(\ref{eq:detected_symbol}) similar to
(\ref{eq:MLdecoding_PPMB_precoded}) and
(\ref{eq:MLdecoding_PPMB_non_precoded}), the bit metrics, equivalent
to (\ref{eq:ML_bit_metrics}) for partial precoding, are
\begin{align}
\gamma^{l,i}(\mathbf{r}_{k}, c_{k'}) =  \left\{
\begin{array}{ll}
\min\limits_{\mathbf{x} \in \psi_{c_{k'}}^{l,i}} \| \mathbf{r}_{k}^p
- \boldsymbol{\Gamma}_p \boldsymbol{\tilde{\Theta}} \mathbf{x} \|^2,
& \textrm{ if $1 \leq l \leq P$} \\
\min\limits_{x \in \chi_{c_{k'}}^{i}} |r_{k,l} - \lambda_{\tilde{l}}
x |^2, & \textrm{ if $P+1 \leq l \leq S$}
\end{array} \right.
\label{eq:ML_bit_metrics_BICMB_CP}
\end{align}
where $\psi_{b}^{l,i}$ is a subset of $\chi^P$, defined as
\begin{equation*}
\psi_{b}^{l,i} = \{ \mathbf{x} = [x_1 \, \cdots \, x_P ]^T :
x_{s|s=l} \in \chi_{b}^{i}, \textrm{ and } x_{s|s \neq l} \in \chi
\},
\end{equation*}
and $\tilde{l}$ is an entry in $\boldsymbol{\omega}$, corresponding
to the subchannel mapped by $\mathbf{T}$. Finally, the ML decoder
makes decisions according to the rule
\begin{align}
\mathbf{\hat{c}} = \arg\min_{\mathbf{\tilde{c}}} \sum_{k'}
\gamma^{l,i}(\mathbf{r}_{k}, \tilde{c}_{k'}).
\label{eq:Decision_Rule}
\end{align}

\section{Diversity Analysis : UMB-CP} \label{sec:div_anal_UMB-CP}

\subsection{Fully Precoded Multiple Beamforming} \label{sec:FPMB}
Based on the ML decoding in (\ref{eq:MLdecoding_PMB}), the upper
bound to the instantaneous PEP between the transmitted symbol
$\mathbf{x}$ and the detected symbol $\mathbf{\hat{x}}$ is
calculated as \ifCLASSOPTIONtwocolumn
\begin{align}
\mathrm{Pr} \left( \mathbf{x} \rightarrow \mathbf{\hat{x}} \mid
\mathbf{H} \right) &= \mathrm{Pr} \left( \left\| \mathbf{y} -
\mathbf{\Gamma} \mathbf{\Theta x} \right\|^2 \geq \left\|\mathbf{y}
- \mathbf{\Gamma} \mathbf{\Theta \hat{x}} \right\|^2 \mid \mathbf{H}
\right) \nonumber \\
&\leq \frac{1}{2} \exp \left(- \frac{\left\| \mathbf{\Gamma}
\mathbf{\Theta} (\mathbf{x} - \mathbf{\hat{x}})\right\|^2}{4 N_0}
\right). \label{eq:instantaneous_PEP_PMB}
\end{align}
\else
\begin{align}
\mathrm{Pr} \left( \mathbf{x} \rightarrow \mathbf{\hat{x}} \mid
\mathbf{H} \right) = \mathrm{Pr} \left( \left\| \mathbf{y} -
\mathbf{\Gamma} \mathbf{\Theta x} \right\|^2 \geq \left\|\mathbf{y}
- \mathbf{\Gamma} \mathbf{\Theta \hat{x}} \right\|^2 \mid \mathbf{H}
\right) \leq \frac{1}{2} \exp \left(- \frac{\left\| \mathbf{\Gamma}
\mathbf{\Theta} (\mathbf{x} - \mathbf{\hat{x}})\right\|^2}{4 N_0}
\right). \label{eq:instantaneous_PEP_PMB}
\end{align}
\fi
Let $\mathbf{d} = \left[d_1 \, \cdots \, d_S \right] ^T =
\mathbf{\Theta} ( \mathbf{x} - \mathbf{\hat{x}} )$. Then, for FPMB,
the average PEP becomes
\begin{align}
\mathrm{Pr} \left( \mathbf{x} \rightarrow \mathbf{\hat{x}} \right)
&\leq E \left[ \frac{1}{2} \exp \left(- \frac{\sum\limits_{s=1}^S
\lambda_s^2 | d_s | ^2}{4 N_0} \right) \right]. \label{eq:PEP_FPMB}
\end{align}
In \cite{ParkICC09}, we showed that equations in the form of
(\ref{eq:PEP_FPMB}) have a closed form upper bound expression. We
provide a formal statement below.
\begin{theorem}
Consider the $S \leq \min(N, M)$ ordered eigenvalues $\mu_1 > \cdots
> \mu_S$ of the uncorrelated central Wishart
matrix\footnote{A central Wishart matrix is the Hermitian matrix
$\mathbf{AA}^H$ where the entry of the matrix $\mathbf{A}$ is
complex Gaussian with zero mean so that $E[ \mathbf{A} ] =
\mathbf{0}$. The Wishart matrix $\mathbf{AA}^H$ is called
uncorrelated if the common covariance matrix, defined as $\mathbf{C}
= E[ \mathbf{a}_s \mathbf{a}_s^H ] \, \forall{s}$, where
$\mathbf{a}_s$ is the $s^{th}$ column vector of $\mathbf{A}$,
satisfies $\mathbf{C} = \mathbf{I}$.} \cite{ZanellaICC08}, and a
weight vector $\boldsymbol{\phi} = [\phi_1 \, \cdots \, \phi_S]^T$
with nonnegative real elements. In the high signal-to-noise ratio
regime, an upper bound for the expression $E [ \exp (-\gamma
\sum_{s=1}^S \phi_s \mu_s ) ]$ which is used in the diversity
analysis of a number of MIMO systems is
\begin{align*}
E\left[ \exp \left( - \gamma \sum\limits_{s=1}^S \phi_s \mu_s
\right) \right] \leq \zeta \left( \phi_{min} \gamma
\right)^{-(N-\delta+1)(M-\delta+1)}
\end{align*}
where $\gamma$ is signal-to-noise ratio, $\zeta$ is a constant,
$\phi_{min} = \min \{ \phi_1, \, \cdots, \, \phi_S \}$, and $\delta$
is the index indicating the first nonzero element in the weight
vector. \label{theorem:E_PEP}
\end{theorem}
\begin{IEEEproof}
See \cite{ParkICC09}.
\end{IEEEproof}
Applying Theorem \ref{theorem:E_PEP} to (\ref{eq:PEP_FPMB}), we get
the upper bound to PEP as
\begin{align}
\mathrm{Pr} \left( \mathbf{x} \rightarrow \mathbf{\hat{x}} \right)
&\leq \tilde{\zeta} \left( \frac{\hat{d}_{min}}{4N} SNR
\right)^{-(N-\delta+1)(M-\delta+1)} \label{eq:PEP_FPMB_final}
\end{align}
where $\tilde{\zeta}$ is a constant, $\hat{d}_{min} = \min
\{|d_1|^2, \, \cdots, \, |d_S|^2\}$, and $\delta$ is an index
indicating the first nonzero element of the vector $\left[ |d_1|^2
\, \cdots \, |d_S|^2 \right]$. Therefore, FPMB achieves the full
diversity order if $\delta$ from any distinct pair is equal to $1$,
which implies that $|d_1|^2 = |\boldsymbol{\theta}_1^T (\mathbf{x} -
\mathbf{\hat{x}})|^2 > 0$ for any distinct pair, where
$\boldsymbol{\theta}_1^T$ is the first row vector of
$\mathbf{\Theta}$. Several methods to build the precoding matrix are
described in \cite{ParkGlobecom09} and \cite{park-2009_arxiv}.

\subsection{Partially Precoded Multiple Beamforming} \label{sec:PPMB}
Generalizing (\ref{eq:instantaneous_PEP_PMB}) for PPMB, we get an
upper bound to PEP as
\begin{align}
\mathrm{Pr} \left( \mathbf{x} \rightarrow \mathbf{\hat{x}} \right)
&\leq E \left[ \frac{1}{2} \exp \left(- \frac{\kappa}{4 N_0} \right)
\right] \label{eq:PEP_PPMB_equation}
\end{align}
where
\begin{align}
\kappa = \sum\limits_{s = 1}^P \lambda_{\eta_s}^2 | \tilde{d}_s | ^2
&+ \sum\limits_{s = 1}^{S-P} \lambda_{\omega_s}^2 | x_{\omega_s} -
\hat{x}_{\omega_s} | ^2 \label{eq:PEP_PPMB}
\end{align}
and $\tilde{d}_s$ is the $s^{th}$ element of a vector
$\mathbf{\tilde{d}} =$ \mbox{$\mathbf{\tilde{\Theta}}
(\mathbf{x}_{\boldsymbol{\eta}} -
\mathbf{\hat{x}}_{\boldsymbol{\eta}})$}. Let us assume that the
constellation precoding matrix $\mathbf{\tilde{\Theta}}$ meets the
condition of FPMB to achieve the full diversity order. Since the
expression (\ref{eq:PEP_PPMB_equation}) with (\ref{eq:PEP_PPMB}) has
a closed form expression similar to (\ref{eq:PEP_FPMB_final}) as
described in FPMB, the $\delta$ value needs to be obtained from a
composite vector with the elements as $|\tilde{d}_s|^2$ and
$|x_{\omega_s} - \hat{x}_{\omega_s}|^2$, to observe the diversity
behavior of a given pairwise error. In addition, a different pair
can lead to different diversity behavior. Therefore, we need to get
the maximum $\delta$ out of all the possible pairwise errors to
decide the diversity order of a given PPMB system.

All of the distinct pairs of $\mathbf{x}$ and $\mathbf{\hat{x}}$ are
divided into three groups in terms of
$\mathbf{x}_{\boldsymbol{\eta}}$,
$\mathbf{\hat{x}}_{\boldsymbol{\eta}}$,
$\mathbf{x}_{\boldsymbol{\omega}}$, and
$\mathbf{\hat{x}}_{\boldsymbol{\omega}}$. The first group includes
the pairs that have \mbox{$\mathbf{x}_{\boldsymbol{\eta}} =
\mathbf{\hat{x}}_{\boldsymbol{\eta}}$} but
\mbox{$\mathbf{x}_{\boldsymbol{\omega}} \neq
\mathbf{\hat{x}}_{\boldsymbol{\omega}}$}, and the second group
comprises the pairs satisfying \mbox{$\mathbf{x}_{\boldsymbol{\eta}}
\neq \mathbf{\hat{x}}_{\boldsymbol{\eta}}$} but
\mbox{$\mathbf{x}_{\boldsymbol{\omega}} =
\mathbf{\hat{x}}_{\boldsymbol{\omega}}$}. Finally, the last group
consists of the pairs for which
\mbox{$\mathbf{x}_{\boldsymbol{\eta}} \neq
\mathbf{\hat{x}}_{\boldsymbol{\eta}}$} and
\mbox{$\mathbf{x}_{\boldsymbol{\omega}} \neq
\mathbf{\hat{x}}_{\boldsymbol{\omega}}$}. We will present the method
to calculate the maximum $\delta$ for each group, and to find
$\delta_{max}$ from the groups.

Since the vector $\mathbf{\tilde{d}}$ is a zero vector for the first
group, the first summation of $\kappa$ in (\ref{eq:PEP_PPMB}) is
zero, resulting in $\delta$ being equal to the minimum of
$\boldsymbol{\omega}$. By considering all of the possible pairs, we
easily see that $\omega_1 \leq \delta \leq \omega_{(S-P)}$.
Therefore, the maximum value is $\delta_1 = \omega_{(S-P)}$ which
corresponds to the pair satisfying $x_s = \hat{x}_s$ for all $s$
except \mbox{$s = \omega_{(S-P)}$}. For any pair in the second
group, the term with the first singular value survives in $\kappa$,
according to the inherited property of the constellation precoding
matrix, i.e., $|\tilde{d}_1|^2 > 0$. However, the second summation
in $\kappa$ disappears since $\mathbf{x}_{\boldsymbol{\omega}} =
\mathbf{\hat{x}}_{\boldsymbol{\omega}}$. Therefore, the maximum
value of this group is $\delta_2 = \eta_1$. Now, for the third
group, both summations in $\kappa$ exist. Then, $\delta$ is chosen
to be the smaller value between the minimum of $\boldsymbol{\omega}$
and $\eta_1$. In the same manner as was already given in the
analysis of the first group, the maximum of the minimum of
$\boldsymbol{\omega}$ is found to be $\omega_{(S-P)}$. Therefore,
the maximum $\delta$ for this group is $\delta_3 = \max \{ \eta_1,
\, \omega_{(S-P)} \}$. Finally, $\delta_{max}$ is decided as
\begin{align}
\delta_{max} &= \max \{ \delta_1, \, \delta_2, \, \delta_3 \} = \max
\left( \eta_1, \, \omega_{(S-P)} \right). \label{eq:Q_max}
\end{align}
\textbf{Example:} We provide the diversity analysis of the $4 \times
4$ PPMB system with \mbox{$S = 4$} and \mbox{$P = 2$}. In this
example, we assume that the precoded symbols are transmitted on the
subchannel $1$ and $3$ while the non-precoded symbols are
transmitted on the subchannel $2$ and $4$. Then, this configuration
gives $\boldsymbol{\eta} = \left[ 1 \,\, 3 \right]$, and
$\boldsymbol{\omega} = \left[ 2 \,\, 4 \right]$. By following the
result in (\ref{eq:Q_max}), $\delta_{max}$ is equal to $\max
\left(1, \, 4 \right) = 4$, leading to the diversity order of $1$.
The pairwise errors, satisfying $x_1 = \hat{x}_1, x_2 = \hat{x}_2,
x_3 = \hat{x}_3$, but $x_4 \neq \hat{x}_4$, inflict loss on the
diversity order of this system. Table \ref{tab:partial_order}
summarizes the diversity order analysis for all of the possible
combinations of the $4 \times 4$ PPMB system. We will provide
simulation results that verify this analysis in Section
\ref{sec:results}, specifically in \figurename{
\ref{fig:partial_precoding}}.

\ifCLASSOPTIONonecolumn
\begin{table}[!m]
\else
\begin{table}[!t]
\fi \caption{Diversity order ($O_{div}$) of $4 \times 4$, $S = 4$
partially precoded multiple beamforming system}
\begin{center}
\begin{tabular}{|c|c|c|c|c|c|c|}
\hline
$P$ & $\boldsymbol{\eta}$ & $\boldsymbol{\omega}$ & $\eta_1$ & $\omega_{(S-P)}$ & $\delta_{max}$ & $O_{div}$\\
\hline \hline
\multirow{6}*{$2$} & $[1 \, 2]$ & $[3 \, 4]$ & $1$ & $4$ & $4$ & $1$ \\
\cline{2-7}
& $[1 \, 3]$ & $[2 \, 4]$ & $1$ & $4$ & $4$ & $1$ \\
\cline{2-7}
& $[1 \, 4]$ & $[2 \, 3]$ & $1$ & $3$ & $3$ & $4$ \\
\cline{2-7}
& $[2 \, 3]$ & $[1 \, 4]$ & $2$ & $4$ & $4$ & $1$ \\
\cline{2-7}
& $[2 \, 4]$ & $[1 \, 3]$ & $2$ & $3$ & $3$ & $4$ \\
\cline{2-7}
& $[3 \, 4]$ & $[1 \, 2]$ & $3$ & $2$ & $3$ & $4$ \\
\hline \hline
\multirow{4}*{$3$} & $[1 \, 2 \, 3]$ & $[4]$ & $1$ & $4$ & $4$ & $1$ \\
\cline{2-7}
& $[1 \, 2 \, 4]$ & $[3]$ & $1$ & $3$ & $3$ & $4$ \\
\cline{2-7}
& $[1 \, 3 \, 4]$ & $[2]$ & $1$ & $2$ & $2$ & $9$ \\
\cline{2-7}
& $[2 \, 3 \, 4]$ & $[1]$ & $2$ & $1$ & $2$ & $9$ \\
\hline
\end{tabular}
\end{center}
\label{tab:partial_order}
\end{table}

\section{Diversity Analysis : BICMB-CP} \label{sec:div_anal_BICMB-CP}

\subsection{BICMB with Full Precoding} \label{sec:BICMB-FP}
We assume that the $d_H$ coded bits are interleaved such that they
are placed in distinct symbols, where $d_H$ denotes the Hamming
distance between the transmitted codeword $\mathbf{c}$ and the
decoded codeword $\mathbf{\hat{c}}$. Since the bit metrics in
(\ref{eq:ML_bit_metrics}) are the same for the same coded bits
between the pairwise errors, the original PEP is replaced by
\begin{align}
\textrm{Pr} \left(\mathbf{c} \rightarrow \mathbf{\hat{c}} |
\mathbf{H}\right) = \textrm{Pr} \left( \sum_{k, d_H}
\min_{\mathbf{x} \in \xi_{c_{k'}}^{l,i}} \| \mathbf{r}_k -
\boldsymbol{\Gamma} \boldsymbol{\Theta} \mathbf{x} \|^2 \geq \right. \ifCLASSOPTIONtwocolumn \nonumber \\
\fi \left. \sum_{k, d_H} \min_{\mathbf{x} \in
\xi_{\hat{c}_{k'}}^{l,i}} \| \mathbf{r}_k - \boldsymbol{\Gamma}
\boldsymbol{\Theta} \mathbf{x} \|^2 \right)
\label{eq:PEP_for_different_codedbits}
\end{align}
where the summation is restricted to the symbols corresponding to
the different $d_H$ coded bits.

Let us define $\mathbf{\tilde{x}}_k$ and $\mathbf{\hat{x}}_k$ as
\begin{equation}
\begin{split}
\mathbf{\tilde{x}}_k = \arg \min_{\mathbf{x} \in \xi_{c_{k'}}^{l,i}}
\| \mathbf{r}_k - \boldsymbol{\Gamma} \boldsymbol{\Theta} \mathbf{x} \|^2 \\
\mathbf{\hat{x}}_k = \arg \min_{\mathbf{x} \in
\xi_{\bar{c}_{k'}}^{l,i}} \| \mathbf{r}_k - \boldsymbol{\Gamma}
\boldsymbol{\Theta} \mathbf{x} \|^2
\end{split}
\label{eq:arg_min}
\end{equation}
where $\bar{c}_{k'}$ is the complement of $c_{k'}$ in binary codes.
It is easily found that $\mathbf{\tilde{x}}_k$ is different from
$\mathbf{\hat{x}}_k$ since the sets that the $l^{th}$ symbols belong
to are disjoint, as can be seen from the definition of
$\xi_{c_{k'}}^{l,i}$. In the same manner, we see that $\mathbf{x}_k$
is different from $\mathbf{\hat{x}}_k$. With $\mathbf{\tilde{x}}_k$
and $\mathbf{\hat{x}}_k$, we get, from
(\ref{eq:PEP_for_different_codedbits}),
\begin{align}
\textrm{Pr} \left( \mathbf{c} \rightarrow \mathbf{\hat{c}} |
\mathbf{H} \right) = \textrm{Pr} \left( \sum_{k, d_H} \|
\mathbf{r}_k - \boldsymbol{\Gamma} \boldsymbol{\Theta}
\mathbf{\tilde{x}}_k
\|^2 \geq \right. \ifCLASSOPTIONtwocolumn \nonumber \\
\fi \left. \sum_{k, d_H} \| \mathbf{r}_k - \boldsymbol{\Gamma}
\boldsymbol{\Theta} \mathbf{\hat{x}}_k \|^2 \right).
\label{eq:alt_expression_PEP_diffbits}
\end{align}
Based on the fact that $\| \mathbf{r}_k - \boldsymbol{\Gamma}
\boldsymbol{\Theta} \mathbf{x}_k \|^2 \geq  \| \mathbf{r}_k -
\boldsymbol{\Gamma} \boldsymbol{\Theta} \mathbf{\tilde{x}}_k \|^2$
and the relation in (\ref{eq:detected_symbol}), equation
(\ref{eq:alt_expression_PEP_diffbits}) is upper-bounded by
\begin{align}
\textrm{Pr} (\mathbf{c} \rightarrow \mathbf{\hat{c}} | \mathbf{H})
\leq \textrm{Pr} \left( \beta \geq \sum_{k, d_H} \|
\boldsymbol{\Gamma} \boldsymbol{\Theta} (\mathbf{x}_k -
\mathbf{\hat{x}}_k) \|^2 \right) \label{eq:PEP_upperbounded}
\end{align}
where
\begin{equation*}
\beta = -\sum_{k, d_H} (\mathbf{x}_k - \mathbf{\hat{x}}_k)^H
\boldsymbol{\Theta}^H \boldsymbol{\Gamma} \mathbf{n}_k +
\mathbf{n}_k^H \boldsymbol{\Gamma} \boldsymbol{\Theta} (\mathbf{x}_k
- \mathbf{\hat{x}}_k).
\end{equation*}
Since $\beta$ is a zero mean Gaussian random variable with variance
$2N_0 \sum_{k, d_H} \|\boldsymbol{\Gamma} \boldsymbol{\Theta}
(\mathbf{x}_k - \mathbf{\hat{x}}_k)\|^2$, the right hand side of
(\ref{eq:PEP_upperbounded}) is replaced by the $Q$ function as
\begin{align}
\textrm{Pr} (\mathbf{c} \rightarrow \mathbf{\hat{c}} | \mathbf{H})
\leq Q\left( \sqrt \frac{\sum\limits_{k, d_H} \| \boldsymbol{\Gamma}
\boldsymbol{\Theta} (\mathbf{x}_k - \mathbf{\hat{x}}_k) \|^2}{2N_0}
\right). \label{eq:PEP_Q}
\end{align}
The numerator in (\ref{eq:PEP_Q}) is rewritten as
\begin{align}
\sum\limits_{k, d_H} \| \boldsymbol{\Gamma} \boldsymbol{\Theta}
(\mathbf{x}_k - \mathbf{\hat{x}}_k) \|^2 = \sum\limits_{s=1}^S
\lambda_s^2 \sum\limits_{k, d_H} |d_{k,s}|^2 \label{eq:Numer_FPMB}
\end{align}
where $d_{k,s}$ is the $s^{th}$ entry of the vector $\mathbf{d}_k =
\boldsymbol{\Theta} (\mathbf{x}_k - \mathbf{\hat{x}}_k)$. Using an
upper bound to the $Q$ function, we calculate the average PEP as
\begin{align}
\textrm{Pr} (\mathbf{c} \rightarrow \mathbf{\hat{c}}) \leq E \left[
\exp \left( - \frac{\sum\limits_{s=1}^S \lambda_s^2 \sum\limits_{k,
d_H} |d_{k,s}|^2}{4 N_0} \right) \right] \label{eq:Avg_PEP_FPMB}.
\end{align}

According to Theorem \ref{theorem:E_PEP}, we can evaluate the
diversity order of a given system by calculating the weight vector
whose $s^{th}$ element is $\sum_{k, d_H} |d_{k,s}|^2$. In
particular, if the constellation precoder is designed such that
\begin{align}
|d_{k,1}|^2 = | \boldsymbol{\theta}_1^T ( \mathbf{x}_k -
\mathbf{\hat{x}_k} ) |^2
> 0, \forall (\mathbf{x}_k, \mathbf{\hat{x}}_k
)\label{eq:condition_full_diversity}
\end{align}
where $\boldsymbol{\theta}_1^T$ is the first row vector of the
precoding matrix $\boldsymbol{\Theta}$, we see that $\sum_{k, d_H}
|d_{k,1}|^2 > 0$, resulting in the full diversity order of $NM$.
Therefore, (\ref{eq:condition_full_diversity}) is a sufficient
condition for the full diversity order of BICMB-FP.

\subsection{BICMB with Partial Precoding} \label{sec:BICMB-PP}
The bit metrics in (\ref{eq:ML_bit_metrics_BICMB_CP}) lead to the
PEP calculation as
\begin{align}
\textrm{Pr} \left(\mathbf{c} \rightarrow \mathbf{\hat{c}} |
\mathbf{H}\right) = \textrm{Pr} \left( \tau_1 \geq \tau_2 \right)
\label{eq:PEP_for_different_codedbits_PPMB_case3}
\end{align}
where
\begin{align*}
\tau_1 &= \sum_{k, d_H^p} \min_{\mathbf{x} \in \psi_{c_{k'}}^{l,i}}
\| \mathbf{r}_{k}^p - \boldsymbol{\Gamma}_p
\boldsymbol{\tilde{\Theta}} \mathbf{x} \|^2 + \sum_{k,d_H^n} \min_{x
\in \chi_{c_{k'}}^{l,i}} | r_{k,l} - \lambda_{\tilde{l}} x |^2
\end{align*}
\begin{align*}
\tau_2 &= \sum_{k, d_H^p} \min_{\mathbf{x} \in
\psi_{\bar{c}_{k'}}^{l,i}} \| \mathbf{r}_{k}^p -
\boldsymbol{\Gamma}_p \boldsymbol{\tilde{\Theta}} \mathbf{x} \|^2 +
\sum_{k, d_H^n} \min_{x \in \chi_{\bar{c}_{k'}}^{l,i}} | r_{k, l} -
\lambda_{\tilde{l}} x |^2
\end{align*}
and $\sum_{k, d_H^p}$, $\sum_{k, d_H^n}$ stand for the summation
over the $d_H^p$ and $d_H^n$ bit metrics, with $d_H^p$ and $d_H^n$
denoting the number of different coded bits between the two pairwise
errors residing on the precoded and the non-precoded subchannels
specified by $\boldsymbol{\eta}$ and $\boldsymbol{\omega}$,
respectively. By using the appropriate system input-output
relations, the PEP is written as
\begin{align}
\textrm{Pr} \left(\mathbf{c} \rightarrow \mathbf{\hat{c}} |
\mathbf{H} \right) = \textrm{Pr} \left( \hat{\beta} \geq
\hat{\kappa} \right) \label{eq:PEP_upperbounded_PPMB3}
\end{align}
where $\hat{\beta} = \beta_p + \beta_n$,
\begin{align*}
\beta_p = -\sum_{k, d_H^p} (\mathbf{x}_{k, \boldsymbol{\eta}} -
\mathbf{\hat{x}}_{k, \boldsymbol{\eta}})^H
\boldsymbol{\tilde{\Theta}}^H \boldsymbol{\Gamma}_p \mathbf{n}_{k}^p
+ \ifCLASSOPTIONtwocolumn \nonumber \\ \fi
\left(\mathbf{n}_{k}^p\right)^H \boldsymbol{\Gamma}_p
\boldsymbol{\tilde{\Theta}} (\mathbf{x}_{k, \boldsymbol{\eta}} -
\mathbf{\hat{x}}_{k, \boldsymbol{\eta}}),
\end{align*}
\begin{align*}
\beta_n = - \sum_{k, d_H^n} \lambda_{\tilde{l}} (x_{k, l} -
\hat{x}_{k,l})^*n_{k,l} + \lambda_{\tilde{l}} (x_{k, l} -
\hat{x}_{k,l}) n_{k,l}^*,
\end{align*}
and
\begin{align*}
\hat{\kappa} = \sum_{k, d_H^p} \| \boldsymbol{\Gamma}_p
\boldsymbol{\tilde{\Theta}} \left( \mathbf{x}_{k, \boldsymbol{\eta}}
- \mathbf{\hat{x}}_{k, \boldsymbol{\eta}} \right) \|^2 + \sum_{k,
d_H^n} | \lambda_{\tilde{l}} \left( x_{k,l} - \hat{x}_{k,l} \right)
|^2.
\end{align*}
Since $\hat{\beta}$ in (\ref{eq:PEP_upperbounded_PPMB3}) is a
Gaussian random variable with zero mean and variance $2N_0
\hat{\kappa}$, the PEP can be expressed in a way similar to
(\ref{eq:PEP_Q}) with the $Q$-function. In addition, if we define
$\sigma$ as
\begin{equation}
\sigma = \sum_{r=1}^P \lambda_{\eta_r}^2 \sum_{k, d_H^p}
|\hat{d}_{k, r}|^2 + d^2_{min}\sum_{r=1}^{S-P} \lambda_{\omega_r}^2
\alpha_{\omega_r} \label{eq:kappa_lowerbound}
\end{equation}
where $\hat{d}_{k,r}$ is the $r^{th}$ entry of the vector
$\mathbf{\hat{d}}_k = \boldsymbol{\tilde{\Theta}} \left(
\mathbf{x}_{k, \boldsymbol{\eta}} - \mathbf{\hat{x}}_{k,
\boldsymbol{\eta}} \right)$, and $\alpha_s$ is the number of times
the $s^{th}$ subchannel is used corresponding to $d_H^n$ bits under
consideration, then we can see that $\sigma \leq \hat{\kappa}$.
Finally, the average PEP is calculated as
\begin{align}
\textrm{Pr} \left(\mathbf{c} \rightarrow \mathbf{\hat{c}} \right)
\leq E \left[ \frac{1}{2} \exp \left( - \frac{\sigma}{4 N_0} \right)
\right]. \label{eq:Avg_PEP_PPMB_case3}
\end{align}

To determine the diversity order from $\sigma$, we need to find the
index indicating the first nonzero element in an ordered composite
vector which consists of $\sum_{k, d_H^p} | \hat{d}_{k,r}|^2$ and
$\alpha_{\omega_r}$ as in Theorem \ref{theorem:E_PEP}. If $d_H^p =
0$, the first summation part of $\sigma$ vanishes. In this case, the
first index is
\begin{equation}
\delta = \min\{  s : \alpha_{s}
> 0 \textrm{ for } s \in \{\omega_1, \, \cdots, \, \omega_{(S-P)}\}\}.
\label{eq:delta_PPMB_positive}
\end{equation}
In the other case of $d_H^p > 0$, we see that $\mathbf{x}_{k,
\boldsymbol{\eta}}$ and $\mathbf{\hat{x}}_{k, \boldsymbol{\eta}}$
are obviously different for the same reason as in the previous
section. If the constellation precoder satisfies the sufficient
condition of (\ref{eq:condition_full_diversity}), the term with
$\lambda_{\eta_1}^2$ always exists in $\sigma$. By considering the
second term of $\sigma$, we get $\delta$ for the case of $d_H^p > 0$
\begin{equation}
\delta = \left\{ \begin{array}{ll}
  \min(\eta_1, \delta') & \textrm{ if $\delta'$ exists,} \\
  \eta_1 & \textrm{ otherwise.}
\end{array} \right.
\label{eq:delta_PPMB_zero}
\end{equation}
where $\delta'$, if it exists, is obtained in the same way as
(\ref{eq:delta_PPMB_positive}). If, in search of $\delta'$, no $s$
satisfying the right hand side of (\ref{eq:delta_PPMB_positive})
exists, we state $\delta'$ does not exist and set $\delta = \eta_1$,
as in (\ref{eq:delta_PPMB_zero}).
\\\textbf{Example:}
In this example, we employ $4$-state $1/2$-rate convolutional code
with generator polynomials $(5, 7)$ in octal representation, in an
$N = M = S = 3$ system. Two types of spatial interleavers are used
to demonstrate the different results of the diversity order. A
generalized transfer function of BICMB with the specific spatial
interleaver and convolutional code provides the $\alpha$-vectors for
all of the pairwise errors, whose element indicates the number of
times the stream is used for the erroneous bits \cite{ParkICC09}. In
particular, due to the fact that $d_H^p = \sum_{r=1}^P
\alpha_{\eta_r}$ and $d_H^n = \sum_{r=1}^{S-P} \alpha_{\omega_r}$
where $\alpha_s$ is the $s^{th}$ element of the $\alpha$-vector, the
generalized transfer function approach in \cite{ParkICC09} is also
useful in the analysis of BICMB-PP. Hence, we rewrite the transfer
functions of the systems from \cite{ParkICC09}, where $a$, $b$, and
$c$ are the symbolic representation of the $1^{st}, 2^{nd}, 3^{rd}$
streams, respectively. The spatial interleaver used in
$\mathcal{T}_1$ is a simple rotating switch on $3$ streams. For
$\mathcal{T}_2$, the $u^{th}$ coded bit is interleaved into the
stream $s_{\mathrm{mod}(u-1, 18)+1}$ where $s_1$ = $\cdots$ = $s_6$
= $1$, $s_7$ = $\cdots$ = $s_{12}$ = $2$, $s_{13}$ = $\cdots$ =
$s_{18}$ = $3$ and $\mathrm{mod}$ is the modulo operation. Each term
represents an $\alpha$-vector, and the powers of $a$, $b$, $c$ in
this term indicate the elements of the $\alpha$-vector corresponding
to that term. \ifCLASSOPTIONonecolumn
\begin{align} \mathcal{T}_1 &= Z^5(a^2 b^2 c + a^2 b c^2 + a b^2
c^2) + Z^6(a^3 b^2 c + a^2 b^3 c + a^3 b c^2 + a b^3 c^2 + a^2 b c^3
+ a b^2 c^3) \nonumber\\
&+ Z^7(2 a^3 b^3 c + 2 a^3 b^2 c^2 + 2 a^2 b^3 c^2 + 2 a^3 b c^3 + 2
a^2 b^2 c^3 + 2 a b^3 c^3) \label{eq:Transfunc_S3R1_2} \\
&+ Z^8(a^5 b^3 + a^4 b^3 c + a^3 b^4 c + 2 a^4 b^2 c^2 + 3 a^3 b^3
c^2 + 2 a^2 b^4 c^2 + a^4 b c^3 + 3 a^3 b^2 c^3 +\nonumber \\
&\quad\qquad 3 a^2 b^3 c^3 + a b^4 c^3 + b^5 c^3 + a^3 b c^4 + 2 a^2
b^2 c^4 + a b^3 c^4 + a^3 c^5) + \cdots \nonumber \\
\mathcal{T}_2 &=Z^5 (a^5 + a^3 b^2 + a^2 b^3 b^5 + a^3 c^2 + b^3 c^2
+ a^2 c^3 + b^2 c^3 + c^5)\nonumber\\
&+ Z^6(a^4 b^2 + 3 a^3 b^3 + a^2 b^4 + a^4 c^2 + 3 a^2 b^2 c^2 + b^4
c^2 + 3 a^3 c^3 + 3 b^3 c^3 + a^2 c^4 + b^2 c^4)
\label{eq:Transfunc_S3R1_2_diffdemux} \\
&+ Z^7(2 a^4 b^3 + 2 a^3 b^4 + a^3 b^3 c + 7 a^3 b^2 c^2 + 7 a^2 b^3
c^2 + 2 a^4 c^3 + a^3 b c^3 + 7 a^2 b^2 c^3 +\nonumber\\
&\quad\qquad a b^3 c^3 + 2 b^4 c^3 + 2 a^3 c^4 + 2 b^3 c^4) + \cdots
\nonumber
\end{align}
\else
\begin{align}\label{eq:Transfunc_S3R1_2}
\mathcal{T}_1 &= Z^5(a^2 b^2 c + a^2 b c^2 + a b^2 c^2) \nonumber\\
&+ Z^6(a^3 b^2 c + a^2 b^3 c + a^3 b c^2 + \nonumber\\
&\quad\qquad a b^3 c^2 + a^2 b c^3 + a b^2 c^3) \nonumber\\
&+ Z^7(2 a^3 b^3 c + 2 a^3 b^2 c^2 + 2 a^2 b^3 c^2 + \nonumber\\
&\quad\qquad 2 a^3 b c^3 + 2 a^2 b^2 c^3 + 2 a b^3 c^3)\\
&+ Z^8(a^5 b^3 + a^4 b^3 c + a^3 b^4 c + 2 a^4 b^2 c^2 +\nonumber\\
&\quad\qquad 3 a^3 b^3 c^2 + 2 a^2 b^4 c^2 + a^4 b c^3 + 3 a^3 b^2 c^3 +\nonumber\\
&\quad\qquad 3 a^2 b^3 c^3 + a b^4 c^3 + b^5 c^3 + a^3 b c^4 + \nonumber\\
&\quad\qquad 2 a^2 b^2 c^4 + a b^3 c^4 + a^3 c^5) + \cdots \nonumber
\end{align}
\begin{align}\label{eq:Transfunc_S3R1_2_diffdemux}
\mathcal{T}_2 &=Z^5 (a^5 + a^3 b^2 + a^2 b^3 +\nonumber\\
&\quad\qquad b^5 + a^3 c^2 + b^3 c^2 + a^2 c^3 + b^2 c^3 + c^5)\nonumber\\
&+ Z^6(a^4 b^2 + 3 a^3 b^3 + a^2 b^4 + a^4 c^2 + 3 a^2 b^2 c^2 +\nonumber\\
&\quad\qquad b^4 c^2 + 3 a^3 c^3 + 3 b^3 c^3 + a^2 c^4 + b^2 c^4) \\
&+ Z^7(2 a^4 b^3 + 2 a^3 b^4 + a^3 b^3 c + 7 a^3 b^2 c^2 +\nonumber\\
&\quad\qquad 7 a^2 b^3 c^2 + 2 a^4 c^3 + a^3 b c^3 + 7 a^2 b^2 c^3 +\nonumber\\
&\quad\qquad a b^3 c^3 + 2 b^4 c^3 + 2 a^3 c^4 + 2 b^3 c^4) + \cdots
\nonumber
\end{align}
\fi

Consider the case $\boldsymbol{\eta} = [1 \, 2]$. We see that all of
the $\alpha$-vectors of $\mathcal{T}_1$ have $d_H^p > 0$. Since
$\eta_1 = 1$, $\delta$ equals $1$ whether $\delta'$ exists or not.
In fact, $\delta'$ does not exist for the term $Z^8a^5b^3$.
Therefore, the $\mathcal{T}_1$ BICMB-PP system with
$\boldsymbol{\eta} = [1 \, 2]$ achieves the full diversity order
while BICMB without constellation precoding \cite{ParkICC09}, or
PPMB without Bit-Interleaved Coded Modulation (BICM) loses the full
diversity order \cite{ParkGlobecom09}, \cite{park-2009_arxiv}. For
$\mathcal{T}_2$, the $\alpha$-vector $[0 \, 0 \, 5]$ gives $d_H^p =
0$, resulting in $\delta = 3$. Therefore, the $\mathcal{T}_2$
BICMB-PP system with $\boldsymbol{\eta} = [1 \, 2]$ does not achieve
the full diversity order.

The same analysis for $\boldsymbol{\eta} = [1 \, 3]$ results in the
diversity order of $9$, and $[2 \, 3]$ results in $4$ for the
transfer function $\mathcal{T}_1$. Similarly, both of $[1 \, 3]$ and
$[2 \, 3]$ result in the diversity of $4$ for $\mathcal{T}_2$. As a
consequence, we find that proper selection of the subchannels for
precoding, as well as the appropriate pattern of the spatial
interleaver, is important to achieve the full diversity order of
BICMB-PP. We will present simulation results that verify this
analysis in Section \ref{sec:results}, in particular, in
\figurename{ \ref{fig:3x3_4Q_BICPPMB}}.

\section{Reduced Computational Complexity Sphere Detection}
\label{sec:sphere}
In this section, we will describe the reduced
computational complexity sphere detection for constellation precoded
multiple beamforming with square QAM modulation. More specifically,
we propose the sphere detection technique to reduce the number of
multiplications without losing the performance. Since detecting the
transmitted non-precoded symbols for UMB-CP in
(\ref{eq:MLdecoding_PPMB_non_precoded}) and finding the bit metrics
of non-precoded symbols for BICMB-CP in
(\ref{eq:ML_bit_metrics_BICMB_CP}) can be carried out independently
of the symbols on the other subchannels, we focus on the precoded
$P$ symbols.

Solving (\ref{eq:MLdecoding_PPMB_precoded}) for the ML detection is
well-known to be NP-hard, given that a full search over the entire
lattice space is performed \cite{HassibiICASSP02}. SD, on the other
hand, solves (\ref{eq:MLdecoding_PPMB_precoded}) by searching only
lattice points that lie inside a sphere of radius $\rho$ centering
around the received vector $\mathbf{y}_p$. A frequently used
solution for the QAM-modulated complex signal model is to decompose
the $P$-dimensional complex-valued problem
(\ref{eq:MLdecoding_PPMB_precoded}) into a $2P$-dimensional
real-valued problem, which is written as \ifCLASSOPTIONonecolumn
\begin{align}
\mathbf{\bar{y}} =
\begin{bmatrix}
\Re \{ \mathbf{y}_p \} \\
\Im \{ \mathbf{y}_p \}
\end{bmatrix}
= \mathbf{\bar{F}} \mathbf{\bar{x}} + \mathbf{\bar{n}} =
\begin{bmatrix}
\Re \{ \mathbf{F} \} & -\Im \{ \mathbf{F}
\}  \\
\Im \{ \mathbf{F} \} & \Re \{ \mathbf{F} \}
\end{bmatrix}
\begin{bmatrix}
\Re \{ \mathbf{x}_{\boldsymbol{\eta}} \} \\
\Im \{ \mathbf{x}_{\boldsymbol{\eta}} \}
\end{bmatrix} +
\begin{bmatrix}
\Re \{ \mathbf{n}_p \} \\
\Im \{ \mathbf{n}_p \}
\end{bmatrix},
\label{eq:conventional_lattice}
\end{align}
\else
\begin{align}
\begin{split}
\mathbf{\bar{y}} &=
\begin{bmatrix}
\Re \{ \mathbf{y}_p \} \\
\Im \{ \mathbf{y}_p \}
\end{bmatrix}
= \mathbf{\bar{F}} \mathbf{\bar{x}} + \mathbf{\bar{n}} \\ &=
\begin{bmatrix}
\Re \{ \mathbf{F} \} & -\Im \{ \mathbf{F}
\}  \\
\Im \{ \mathbf{F} \} & \Re \{ \mathbf{F} \}
\end{bmatrix}
\begin{bmatrix}
\Re \{ \mathbf{x}_{\boldsymbol{\eta}} \} \\
\Im \{ \mathbf{x}_{\boldsymbol{\eta}} \}
\end{bmatrix}
+
\begin{bmatrix}
\Re \{ \mathbf{n}_p \} \\
\Im \{ \mathbf{n}_p \}
\end{bmatrix}
\label{eq:conventional_lattice}
\end{split}
\end{align}
\fi where $\mathbf{F} = \boldsymbol{\Gamma}_p
\boldsymbol{\tilde{\Theta}}$ \cite{JaldenJSP05},
\cite{HassibiICASSP02}. The QR decomposition of the $2P \times 2P$
real-valued channel matrix turns (\ref{eq:MLdecoding_PPMB_precoded})
into the equivalent expression
\begin{align}
\mathbf{\hat{x}}_{\boldsymbol{\eta}} = \arg \min_{\mathbf{x} \in
\Psi} \left\| \mathbf{\bar{Q}}^H \mathbf{\bar{y}} - \mathbf{\bar{R}}
\mathbf{x} \right\|^2 \label{eq:QR_Metric}
\end{align}
where $\mathbf{\bar{Q}}$ and $\mathbf{\bar{R}}$ are the unitary
matrix and the upper triangular matrix from the QR decomposition of
$\mathbf{\bar{F}}$ \cite{JaldenJSP05}, \cite{HassibiICASSP02}. Let
$\Omega$ denote the set of scalar symbols for one dimension of QAM,
e.g., $\Omega=\{ -3,-1,1,3\}$ for $16$-QAM, then $\Psi$ denotes a
subset of $\Omega^{2P}$ whose elements satisfy $\Vert
\mathbf{\bar{Q}}^H \mathbf{\bar{y}}-\mathbf{\bar{R}} \mathbf{x}
\Vert^{2}<\rho^{2}$. The initial radius $\rho$ should be chosen
properly so that it is neither too small nor too large. Too small an
initial radius can result in too many unsuccessful searches by
restarting the search and thus increasing the complexity, while too
large an initial radius can result in too many lattice points to be
searched.

The SD algorithm can be viewed as a pruning algorithm on a tree of
depth $2P$, whose branches correspond to elements drawn from the set
$\Omega$ \cite{AzzamGLOBECOM07}, \cite{HassibiICASSP02}.
Conventional SD implements a Depth-First Search (DFS) strategy in
the tree which achieves ML performance. The complexity of SD is
measured in terms of the number of operations required per visited
node multiplied by the number of visited nodes throughout the search
algorithm \cite{HassibiICASSP02}. The complexity can be reduced by
either reducing the number of nodes to be visited or the number of
operations to be carried out at each node or both. In order to
reduce the number of visited nodes, one can either make a judicious
choice of the initial radius to start the algorithm, or execute a
proper sphere radius update strategy. The former strategy has been
studied in \cite{HanGLOBECOM05} and \cite{ChengISCC07}, and the
latter one has been discussed in \cite{HassibiJSP05} and
\cite{ZhaoJCOM05}. In this paper, we propose methods to reduce the
average number of real multiplications, which are the most expensive
operations in terms of machine cycles required at each node for
conventional SD. A proper choice of the initial radius for BICMB-CP
will also be provided.

We start by writing the node weight as \cite{AzzamGLOBECOM07}
\begin{equation}
w(\mathbf{\bar{x}}^{(u)})=w(\mathbf{\bar{x}}^{(u+1)})+w_{pw}(\mathbf{\bar{x}}^{(u)})
\label{eq:nodeweight}
\end{equation}
with $u=2P, \, 2P-1, \, \cdots, \, 1$,
$w(\mathbf{\bar{x}}^{(2P+1)})=0$, and
$w_{pw}(\mathbf{\bar{x}}^{(2P+1)})=0$, where
$\mathbf{\bar{x}}^{(u)}$ denotes the partial vector symbol at layer
$u$. The partial weight $w(\mathbf{\bar{x}}^{(u)})$ is written as
\begin{equation}
w_{pw}(\mathbf{\bar{x}}^{(u)})=|\tilde{y}_u-\sum^{2P}_{v=u}{\bar{R}_{u,v}
\bar{x}_v}|^{2} \label{eq:partial_weight}
\end{equation}
where $\tilde{y}_u$ is the $u^{th}$ element of $\mathbf{\bar{Q}}^H
\mathbf{\bar{y}}$, $\bar{R}_{u,v}$ is the $(u,v)^{th}$ element of
$\mathbf{\bar{R}}$, and $\bar{x}_v$ is the $v^{th}$ element of
$\mathbf{\bar{x}}$.

\subsection{Precalculation of Multiplications}
\label{sec:precalculation}
Note that for one channel realization,
both $\mathbf{\bar{R}}$ and $\Omega$ are independent of time. In
other words, to decode different received symbols for one channel
realization, the only term in (\ref{eq:partial_weight}) which
depends on time is $\tilde{y}_u$. Consequently, a table $\mathbb{T}$
can be constructed to store all terms of $\bar{R}_{u,v} \bar{x}$,
where $\bar{R}_{u,v} \neq 0$ and $\bar{x} \in \Omega$, before
starting the tree search procedure. Equations (\ref{eq:nodeweight})
and (\ref{eq:partial_weight}) imply that only one real
multiplication is needed by using $\mathbb{T}$ instead of $2P-u+2$
for each node to calculate the node weight. As a result, the number
of real multiplications can be significantly reduced.

Taking the square QAM structure into consideration, $\Omega$ can be
divided into two smaller sets $\Omega_{1}$ with negative elements
and $\Omega_{2}$ with positive elements. Take 16-QAM for example,
$\Omega=\{ -3,-1,1,3\}$, then $\Omega_{1}=\{ -3,-1\}$ and
$\Omega_{2}=\{ 1,3\}$. Any negative element in $\Omega_{1}$ has a
positive element with the same absolute value in $\Omega_{2}$.
Consequently, in order to build $\mathbb{T}$, only terms of
$\bar{R}_{u,v} \bar{x}$, where $\bar{R}_{u,v} \neq 0$ and $\bar{x}
\in \Omega_{1}$, need to be calculated and stored. Hence, the size
of $\mathbb{T}$ is
\begin{equation}
|\mathbb{T}|=\frac{N_R|\Omega|}{2} \label{eq:size_T}
\end{equation}
where $N_R$ denotes the number of nonzero elements in matrix
$\mathbf{\bar{R}}$, and $|\Omega|$ denotes the size of $\Omega$.

In order to build $\mathbb{T}$, both the number of terms that need
to be stored and the number of real multiplications required are
$|\mathbb{T}|$. Since the channel is assumed to be flat fading,
only one $\mathbb{T}$ needs to be built in one burst. If the burst
length is very long, the computational complexity of building
$\mathbb{T}$ can be neglected.

\subsection{Modified {DFS} algorithm} \label{sec:mod_dfs_algo}
The representation proposed in \cite{AzzamGLOBECOM07} replaces the
conventional representation of (\ref{eq:conventional_lattice}) with
\begin{align}
\mathbf{\check{y}} = \mathbf{G} \mathbf{\check{x}} +
\mathbf{\check{n}} \label{eq:new_lattice}
\end{align}
where
\begin{equation*}
\mathbf{\check{y}} =
\begin{bmatrix}
\Re \{ y_1 \} & \Im \{ y_1 \} & \cdots & \Re \{ y_P \} & \Im \{ y_P
\}
\end{bmatrix}^{T},
\end{equation*}
\begin{equation*}
\mathbf{G} =
\begin{bmatrix} \Re \{ F_{1,1} \} & -\Im
\{ F_{1,1} \} & \cdots & \Re \{
F_{1,P} \} & -\Im \{ F_{1,P} \} \\
\Im \{ F_{1,1} \} & \Re \{ F_{1,1} \} & \cdots
& \Im \{ F_{1,P} \} & \Re \{ F_{1,P} \} \\
\vdots &  \vdots & \ddots & \vdots & \vdots \\ \Re \{ F_{P,1} \} &
-\Im \{ F_{P,1} \} & \cdots & \Re \{
F_{P,P} \} & -\Im \{ F_{P,P} \} \\
\Im \{ F_{P,1} \} & \Re \{ F_{P,1} \} & \cdots
& \Im \{ F_{P,P} \} & \Re \{ F_{P,P} \} \\
\end{bmatrix},
\end{equation*}
\begin{equation*}
\mathbf{\check{x}} =
\begin{bmatrix}
\Re \{ x_{\eta_1} \} & \Im \{ x_{\eta_1} \} & \cdots & \Re \{
x_{\eta_P} \} & \Im \{ x_{\eta_P} \}
\end{bmatrix}^{T},
\end{equation*}
\begin{equation*}
\mathbf{\check{n}} =
\begin{bmatrix}
\Re \{ n_1 \} & \Im \{ n_1 \} & \cdots & \Re \{ n_P \} & \Im \{ n_P
\}
\end{bmatrix}^{T}.
\end{equation*}

The structure of the lattice representation becomes advantageous
after applying the QR decomposition to $\mathbf{G}$, i.e.,
$\mathbf{G} = \mathbf{Q} \mathbf{R}$. Due to a special form of
orthogonality between each pair of columns, all elements $R_{u,u+1}$
for $u=1, \, 3, \, \cdots, \, 2P-1$, in the upper triangular matrix
$\mathbf{R}$ become zero \cite{AzzamGLOBECOM07}. The locations of
these zeros introduce orthogonality between the real and imaginary
parts of every detected symbol, which can be taken advantage of to
reduce the computational complexity of SD. We provide the following
example to explain this.

Consider a $2 \times 2$ $S = 2$ FPMB system employing $4$-QAM. Then,
SD constructs a tree with $2P=4$ levels, where the branches coming
out of each node represent the real values in the set
$\Omega=\{-1,1\}$. This tree is shown in \figurename{
\ref{fig:tree_structure}}. Based on the representation in
(\ref{eq:new_lattice}), the input-output relation is given by
\begin{equation}
\begin{bmatrix}
\hat{y}_1 \\ \hat{y}_2 \\ \hat{y}_3 \\ \hat{y}_4
\end{bmatrix}
=
\begin{bmatrix}
R_{1,1} & 0 & R_{1,3} & R_{1,4} \\
0 & R_{2,2} & R_{2,3} & R_{2,4} \\
0 & 0 & R_{3,3} & 0 \\
0 & 0 & 0 & R_{4,4}
\end{bmatrix}
\begin{bmatrix}
\check{x}_{1} \\ \check{x}_{2} \\ \check{x}_{3} \\ \check{x}_{4}
\end{bmatrix}
+
\begin{bmatrix}
\hat{n}_{1} \\ \hat{n}_{2} \\ \hat{n}_{3} \\ \hat{n}_{4}
\end{bmatrix}
\label{eq:Example_SD}
\end{equation}
where $\hat{y}_u, \, \check{x}_u, \, \hat{n}_u$ are the $u^{th}$
element of the vectors $\mathbf{Q}^H \mathbf{\check{y}}, \,
\mathbf{\check{x}}, \, \mathbf{Q}^H \mathbf{\check{n}}$,
respectively, and $R_{u, v}$ is the element of $\mathbf{R}$.

\ifCLASSOPTIONonecolumn
\begin{figure}[!m]
\centering \includegraphics[width = 0.6\linewidth]{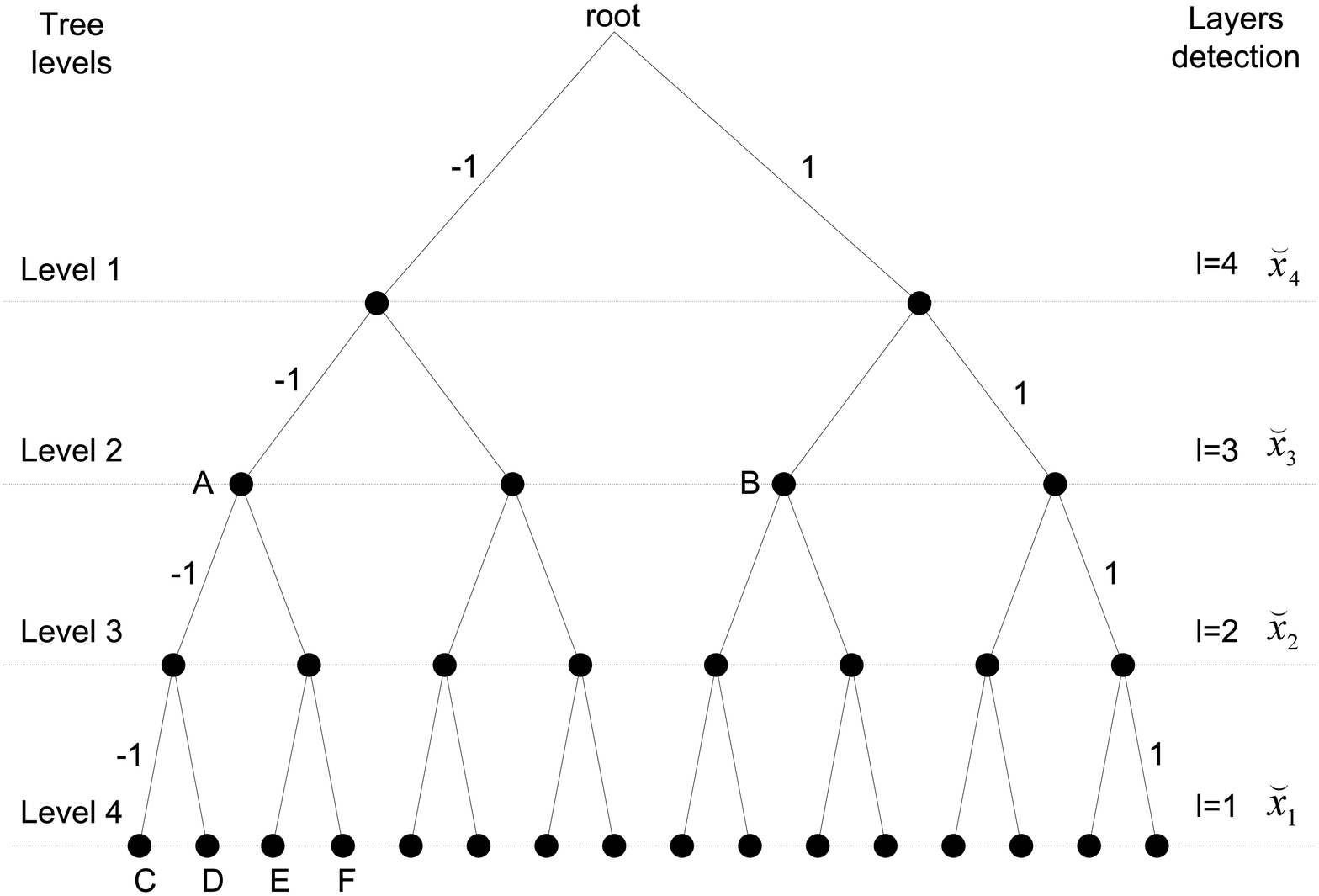}
\caption{Tree structure for a $2 \times 2$ FPMB system employing
$4$-QAM.} \label{fig:tree_structure}
\end{figure}
\else
\begin{figure}[!t]
\centering \includegraphics[width = \sizefig\linewidth]{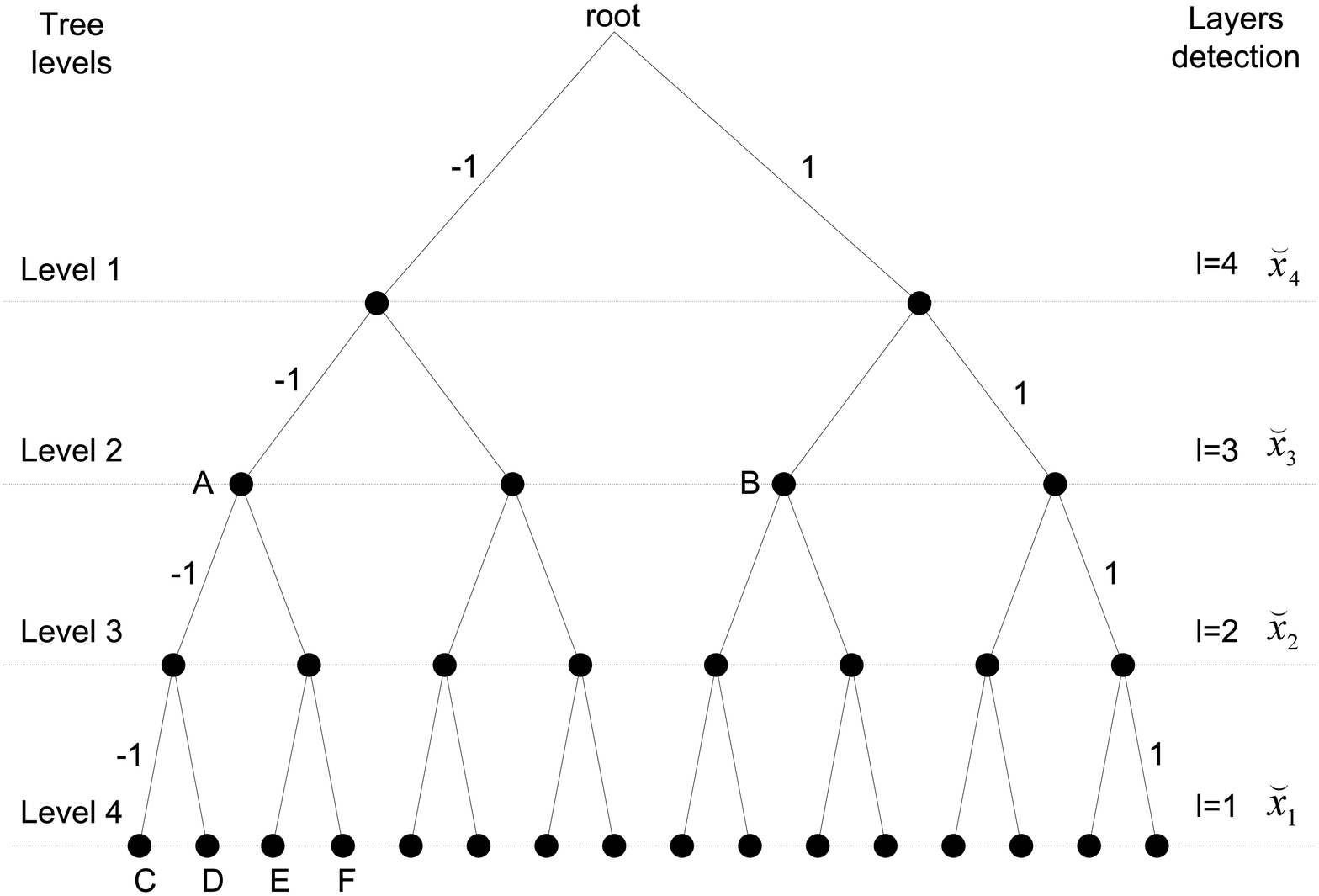}
\caption{Tree structure for a $2 \times 2$ FPMB system employing
$4$-QAM.} \label{fig:tree_structure}
\end{figure}
\fi

Calculating partial node weights of (\ref{eq:Example_SD}) for the
first level and the second level are independent, same as the third
level and the fourth level, because of the additional zeros in the
$\mathbf{R}$ matrix. For instance, the partial weights of node $A$
and $B$ in \figurename{ \ref{fig:tree_structure}} depend on only
$\check{x}_3$, and the partial weights of node $C$, $D$, $E$, and
$F$ depend on $\check{x}_4$, $\check{x}_3$, and $\check{x}_1$ except
$\check{x}_2$. In other words, the partial weights of node $A$ and
$B$ are equal, and need to be calculated once. Similarly, partial
weights of node $C$ and $D$ can be used without an additional
computation for the partial weights of node $E$ and $F$,
respectively.

Because of this feature, the DFS strategy is modified in the
following way: for the $u^{th}$ layer, where $u$ is an odd number,
partial weights of the nodes at the layer $u$ (called \emph{children
nodes}) belonging to a node at the layer $u + 1$ (called \emph{a
parent node}) are stored, and are used as partial weights of the
nodes belonging to the same node at the layer $u+2$ (called \emph{a
grandparent node}), but to the different parent nodes. In other
words, the weights of children nodes belonging to one of the parent
nodes are recycled by the children's \emph{cousins}.

By implementing the modified DFS algorithm, further complexity
reduction is achieved beyond the reduction due to the precalculation
table $\mathbb{T}$. We will show how many real multiplications are
reduced to calculate all nodes at layers $u, u+1$ belonging to one
grandparent node at layer $u+2$, where $u$ is an odd number. Let us
define $\nu \in [0,|\Omega|]$ as the number of non-pruned branches
from the grandparent node, after calculating the node weights
$\omega(\mathbf{\check{x}}^{(u+1)})$ and comparing them with
$\rho^2$. If $\nu = 0$, which means all branches from the
grandparent node are pruned, the modified algorithm does not reduce
computations from the original DFS algorithm. If $\nu > 0$, to get
all of the weights at the layer $u$ and $u+1$ under the grandparent
node, the number of real multiplications reduces further from $(\nu
+ 1)|\Omega|$ to $2|\Omega|$.

\subsection{Initial Radius for BICMB-CP} \label{sec:Initial_Radius}
The proposed SD algorithm for UMB-CP described in the previous
sections can also be applied to BICMB-CP. The $P$-dimensional
complex-valued input-output relation of the precoded part in
(\ref{eq:ML_bit_metrics_BICMB_CP}) can be transformed into a
$2P$-dimensional real-valued problem, based on the lattice
representation in (\ref{eq:new_lattice}). Applying the QR
decomposition to the $2P \times 2P$ dimensional matrix $\mathbf{G}$
in (\ref{eq:new_lattice}), the bit metrics of the precoded part in
(\ref{eq:ML_bit_metrics_BICMB_CP}) are rewritten as
\begin{align}
\gamma^{l,i}(\mathbf{r}_{k}, c_{k'}) = \min_{\mathbf{x} \in
\Phi_{c_{k'}}} \| \mathbf{\hat{r}}_{k} - \mathbf{R} \mathbf{x} \|^2
\label{eq:DFS_bit_metrics}
\end{align}
where $\mathbf{\hat{r}}_{k}$ is the product of $\mathbf{Q}^H$ and
the transformed vector from $\mathbf{r}_k^p$. Due to the
transformation, the position of $c_{k'}$ in the label of
$\mathbf{x}$ needs to be acquired and stored in a new table $k'
\rightarrow (k, \hat{l}, \hat{i})$, which means $c_{k'}$ lies in the
$\hat{i}^{th}$ bit position of label for the $\hat{l}^{th}$ element
of real-valued symbol vector $\mathbf{x}$. Let $\Omega_b^{\hat{i}}$
denote a subset of $\Omega$ whose labels have $b \in \{0, 1\}$ in
the $\hat{i}^{th}$ bit position. If we define
$\tilde{\xi}_{b}^{\hat{l},\hat{i}}$ as
\begin{align*}
\tilde{\xi}_{b}^{\hat{l},\hat{i}} = \{ \mathbf{x} : x_{s|s=\hat{l}}
\in \Omega_{b}^{\hat{i}}, \textrm{ and } x_{s|s \neq \hat{l}} \in
\Omega \}
\end{align*}
then, $\Phi_{b}$ denotes a subset of
$\tilde{\xi}_{b}^{\hat{l},\hat{i}}$, whose elements satisfy $\|
\mathbf{\hat{r}}_{k} - \mathbf{R} \mathbf{x} \|^2 \leq \rho_{b}^2$.

Similarly to UMB-CP, the SD algorithm for BICMB-CP now can be viewed
as a pruning algorithm on a tree of depth $2P$. However, its
branches of the layer $u=\hat{l}$ correspond to elements drawn only
from the set $\chi^{\hat{i}}_{c_{k'}} \subset \chi$. To determine
the initial radius for BICMB-CP, we use the ZF-DFE algorithm to
acquire an estimated real-valued vector symbol $\mathbf{x}_k^{b}$
for $b = 0$ or $1$, whose $u^{th}$ element $x_{k,u}^{b}$ is detected
successively from $x_{k,2P}^{b}$ to $x_{k,1}^{b}$ as
\begin{align}
x_{k,u}^{b} = \arg \min\limits_{x \in \Omega_{b}^{\hat{i}}} |
\hat{r}_{k, u} - \sum\limits_{v=u+1}^{2P} R_{u,v} x_{k,v}^{b} -
R_{u,u}x | \label{eq:estimation_1}
\end{align}
for the element corresponding to $\hat{l}$ indicated by the table
$k' \rightarrow (k, \hat{l}, \hat{i})$, and
\begin{align}
x_{k,u}^{b} = \arg \min\limits_{x \in \Omega} | \hat{r}_{k, u} -
\sum\limits_{v=u+1}^{2P} R_{u,v} x_{k,v}^{b} - R_{u,u}x |
\label{eq:estimation_2}
\end{align}
for the rest of the elements. Then, the initial radius is calculated
by
\begin{align}
\rho_{b}^2 = \| \mathbf{\hat{r}}_{k} - \mathbf{R} \mathbf{x}_k^{b}
\|^2. \label{eq:Initial_radius}
\end{align}
With the initial radius acquired by the ZF-DFE algorithm, the SD
guarantees no unsuccessful search for both of the bit metrics.

\section{Simulation Results} \label{sec:results}

\subsection{UMB-CP} \label{sec:results_uncoded}
To illustrate the analysis of the diversity order in Section
\ref{sec:div_anal_UMB-CP}, we now present simulation results over a
number of different system configurations. \figurename{
\ref{fig:SB_FPMB}} shows BER performance for SB and FPMB. The curves
with the legend FPMB are generated by the precoding matrices that
outperform the others in \cite{ParkGlobecom09},
\cite{park-2009_arxiv}. All of the FPMB systems employ $4$-QAM
modulation, and the system data rate for SB and FPMB is set to $4$,
$8$ bits/channel use for a $2 \times 2$ and a $4 \times 4$ system,
respectively. All of the FPMB systems are shown to achieve the full
diversity order since each slope is parallel to the corresponding SB
system, known to achieve the full diversity order of $NM$.

\ifCLASSOPTIONonecolumn
\begin{figure}[!m]
\centering \includegraphics[width = 0.6\linewidth]{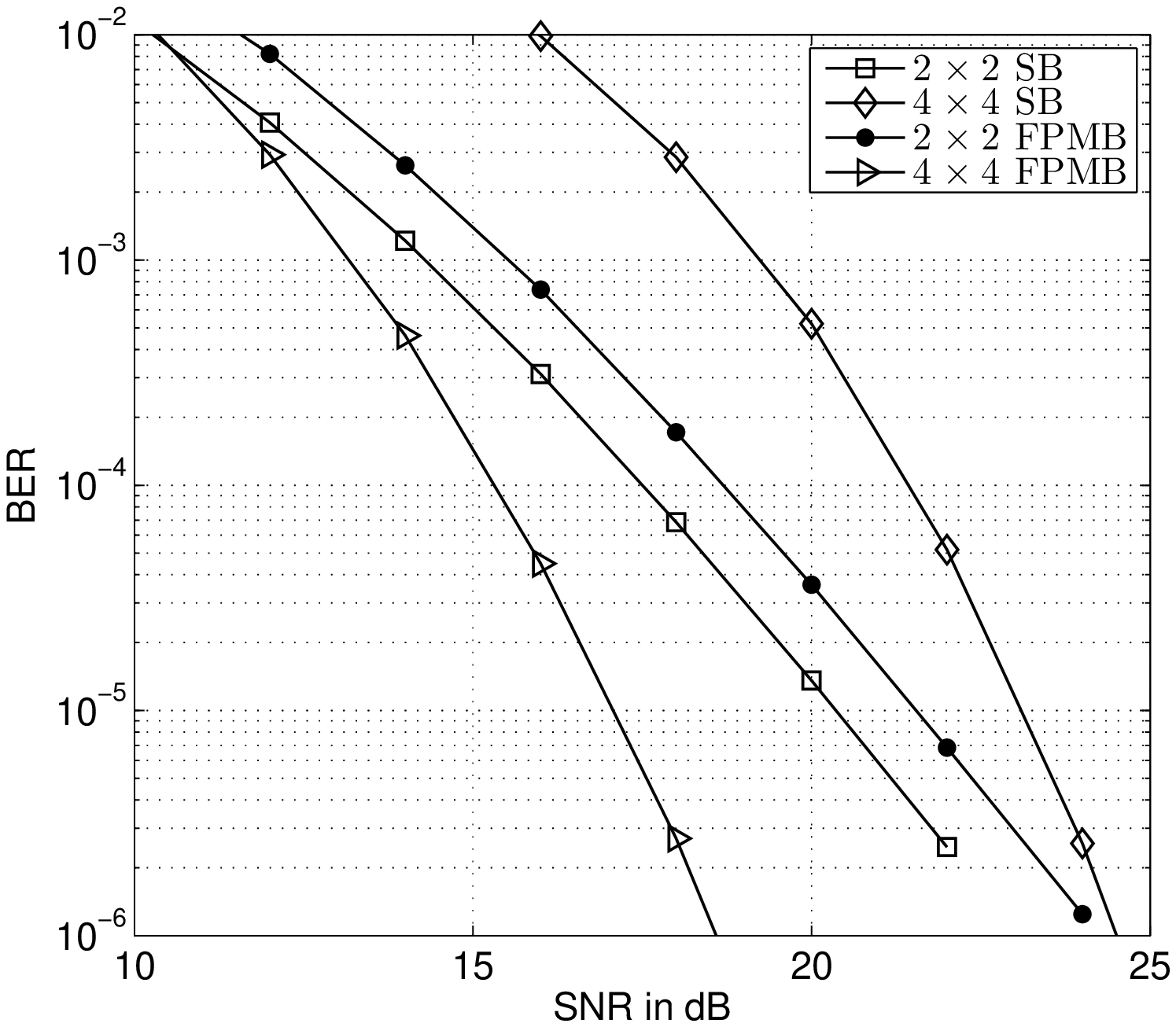}
\caption{BER vs. SNR comparison for $2 \times 2$, $4 \times 4$ SB
and FPMB.} \label{fig:SB_FPMB}
\end{figure}
\else
\begin{figure}[!t]
\centering \includegraphics[width = \sizefig\linewidth]{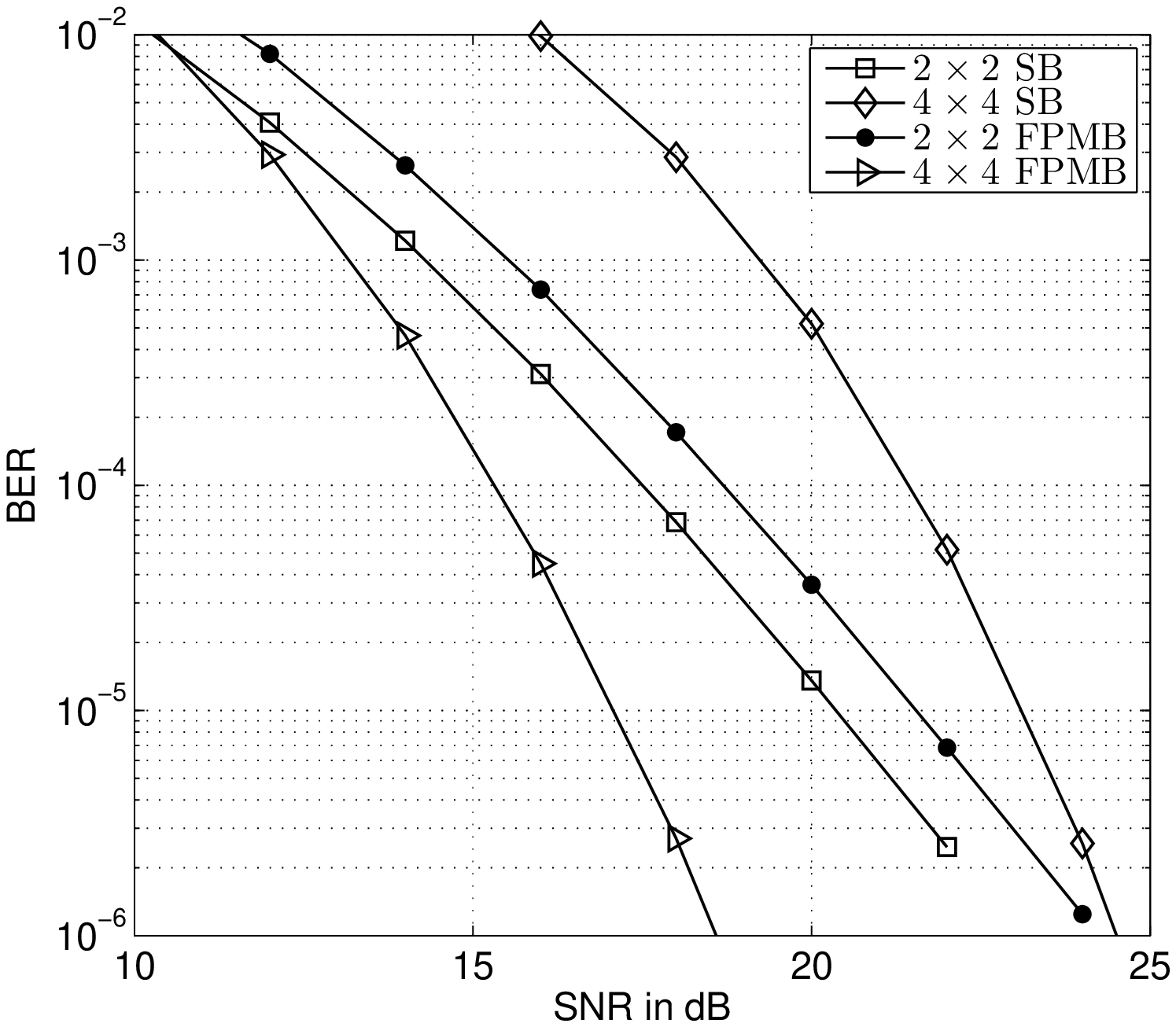}
\caption{BER vs. SNR comparison for $2 \times 2$, $4 \times 4$ SB
and FPMB.} \label{fig:SB_FPMB}
\end{figure}
\fi

Simulation results to support the diversity analysis of $4 \times 4$
$S=4$ PPMB in Table \ref{tab:partial_order} are provided in
\figurename{ \ref{fig:partial_precoding}}. We find that the
simulation results are the same as the diversity orders in Table
\ref{tab:partial_order}.

\ifCLASSOPTIONonecolumn
\begin{figure}[!m]
\centering \includegraphics[width =
0.6\linewidth]{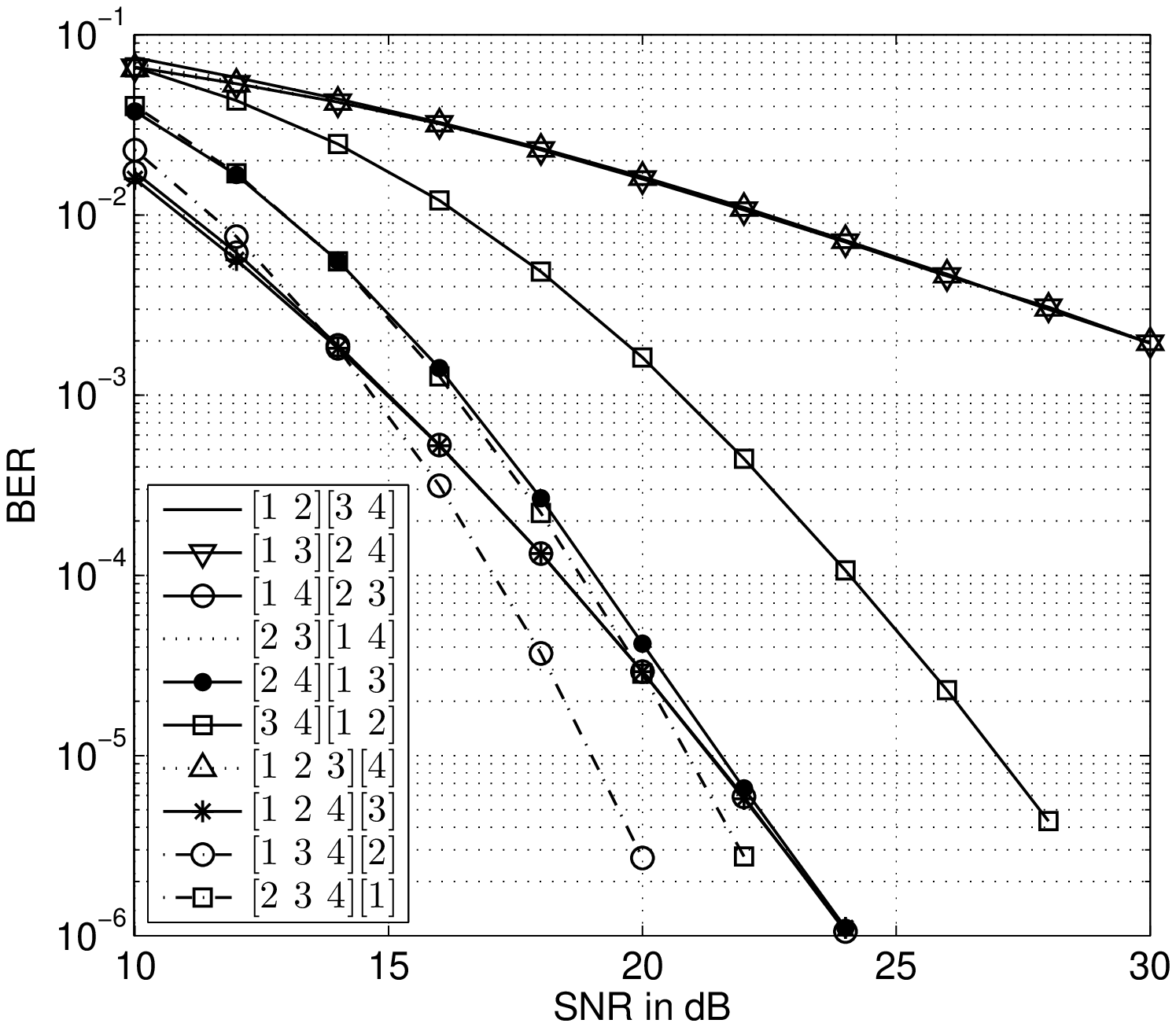} \caption{BER vs. SNR for $4
\times 4$ $S = 4$, $4$-QAM PPMB.} \label{fig:partial_precoding}
\end{figure}
\else
\begin{figure}[!t]
\centering \includegraphics[width =
\sizefig\linewidth]{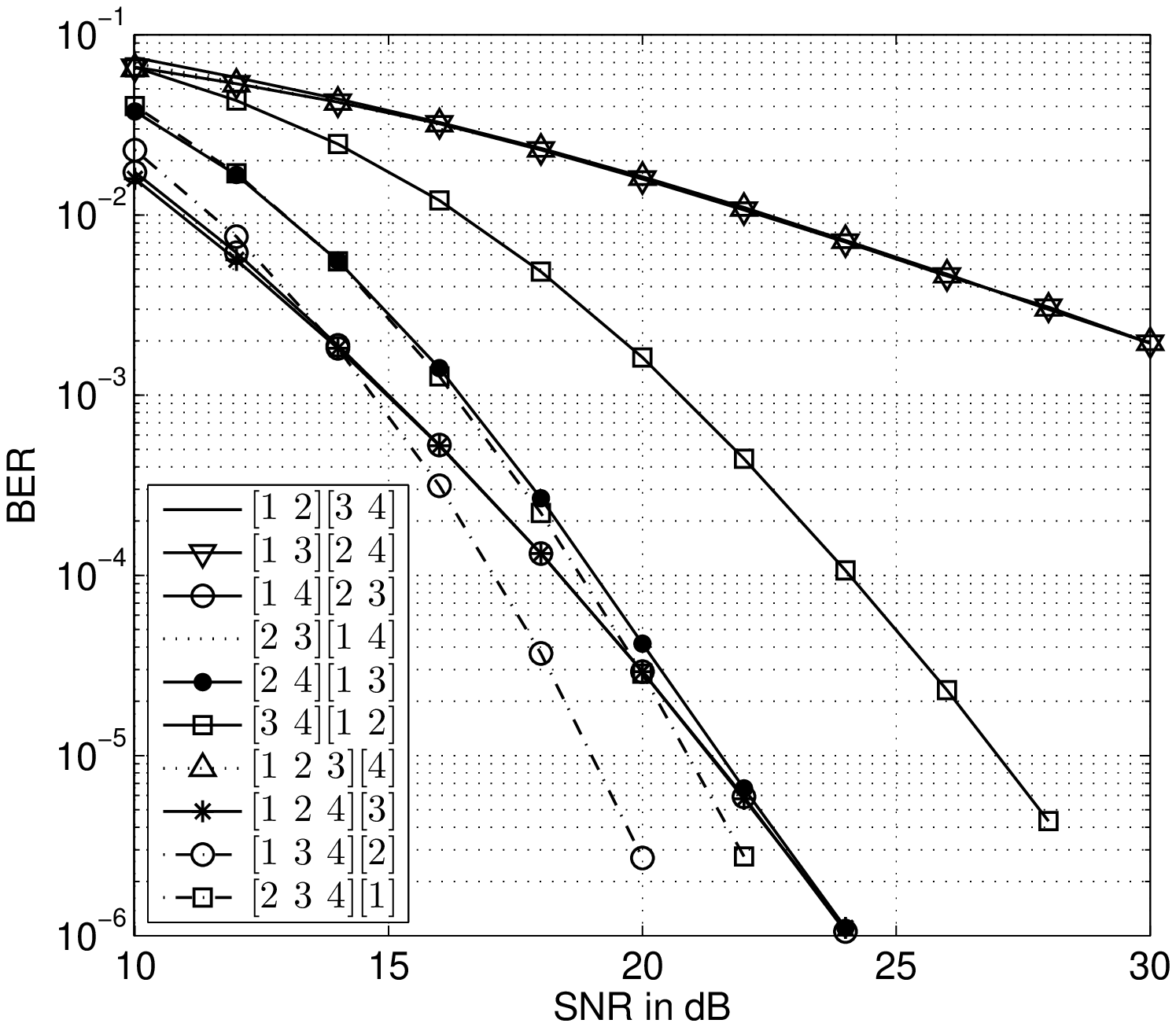} \caption{BER vs. SNR for $4
\times 4$ $S = 4$, $4$-QAM PPMB.} \label{fig:partial_precoding}
\end{figure}
\fi

To verify the reduced computational complexity with sphere detection
in Section \ref{sec:sphere}, we simulated $2 \times 2$ $S=2$ and $4
\times 4$ $S=4$ FPMB systems using $4$-QAM and $64$-QAM with
receivers employing the exhaustive search (EXH), the conventional SD
(CSD), and the proposed SD (PSD). In these simulations, the initial
radius is chosen to be $\rho^2=2N_0P$, inside which at least one
lattice point lies with a high probability \cite{HassibiJSP05}. The
average number of real multiplications for decoding one transmitted
vector symbol is calculated at different SNR. Since the reductions
in complexity are substantial, we will express them as orders of
magnitude (in approximate terms) in the sequel. \figurename{
\ref{fig:mul_2x2_umb}} shows a comparison for the $2 \times 2$ $S=2$
FPMB system. For $4$-QAM, a comparison with EXH shows that CSD
reduces the number of multiplications by approximately $0.6$ and
$0.8$ orders of magnitude at low and high SNR, respectively, and PSD
reduces by approximately $1.0$ and $1.1$ order of magnitude at low
and high SNR, respectively. As seen from the case of $64$-QAM in
\figurename{ \ref{fig:mul_2x2_umb}}, the reduction in complexity
increases as the constellation size increases: the number of
multiplications of CSD decreases by approximately $1.4$ orders of
magnitude at low SNR, and $2.8$ at high SNR, while that of PSD
decreases by $2.4$ and $3.2$ orders of magnitude at low and high
SNR, respectively. \figurename{ \ref{fig:mul_4x4_umb}} shows the
simulation results of $4 \times 4$ $S=4$ FPMB system. For $4$-QAM,
the number of multiplications of CSD is reduced by $1.4$ and $2.1$
orders of magnitude at low and high SNR, respectively. PSD reduces
the complexity by $2.1$ orders of magnitude at low SNR, and $2.4$ at
high SNR. As already observed in \figurename{
\ref{fig:mul_2x2_umb}}, the reduction becomes larger as the
constellation size increases in the $4 \times 4$ $S=4$ FPMB system.
For $64$-QAM, the number of multiplications of CSD decreases by
$3.3$ and $6.4$ orders of magnitude at low and high SNR,
respectively. PSD gives a larger reduction by $4.3$ orders of
magnitude at low SNR, and $7.0$ at high SNR. Simulation results
clearly show that CSD reduces the complexity substantially compared
with EXH, and the complexity can be further reduced effectively by
our PSD. The complexity reduction becomes larger as the
constellation precoder dimension or the constellation size becomes
larger.

\ifCLASSOPTIONonecolumn
\begin{figure}[!m]
\centering \includegraphics[width = 0.6\linewidth]{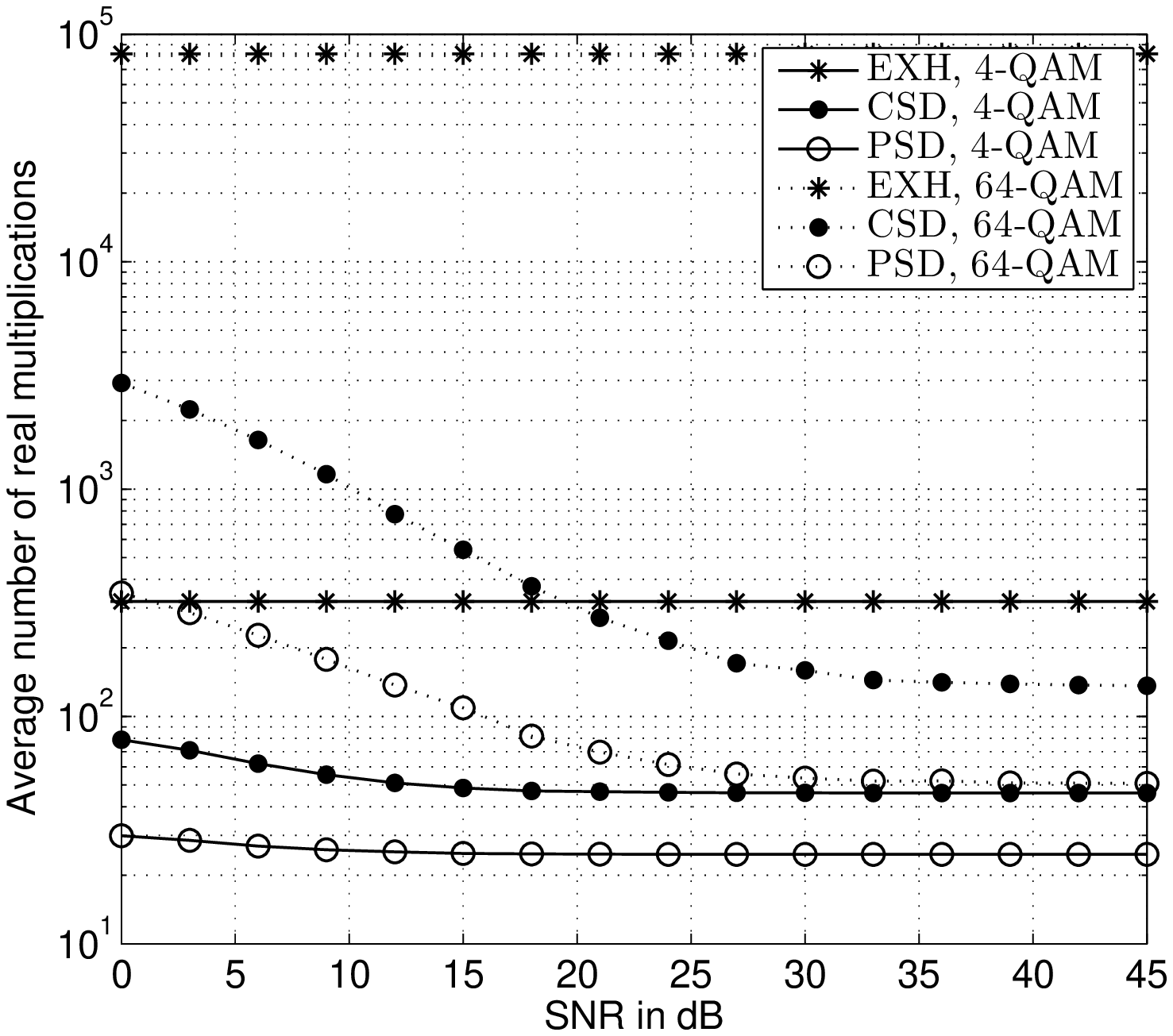}
\caption{Average number of real multiplications vs. SNR for the $2
\times 2$ FPMB systems with $4$-QAM and $64$-QAM.}
\label{fig:mul_2x2_umb}
\end{figure}

\begin{figure}[!m]
\centering \includegraphics[width = 0.6\linewidth]{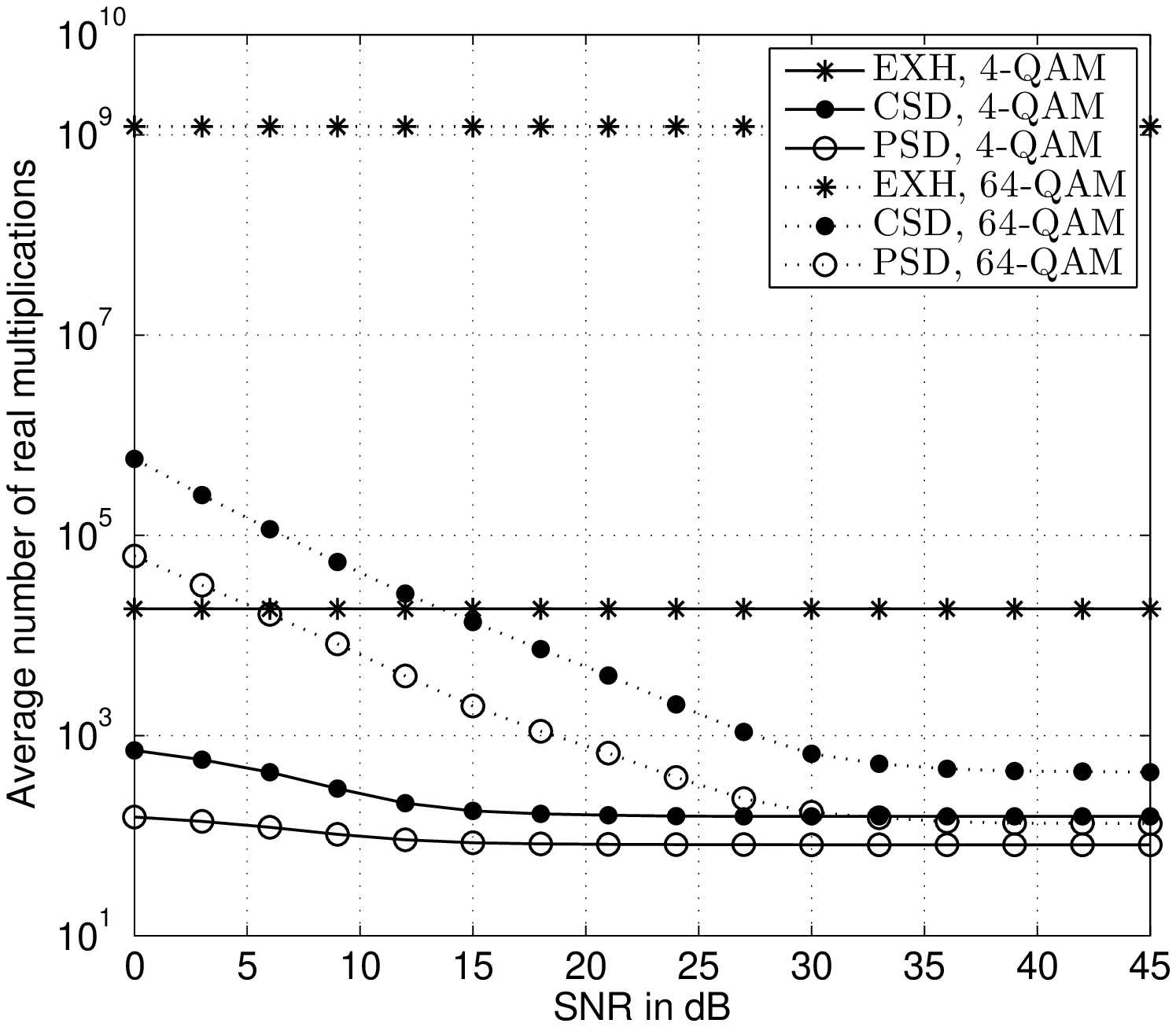}
\caption{Average number of real multiplications vs. SNR for the $4
\times 4$ FPMB systems with $4$-QAM and $64$-QAM.}
\label{fig:mul_4x4_umb}
\end{figure}

\else

\begin{figure}[!t]
\includegraphics[width =
\sizefig\linewidth]{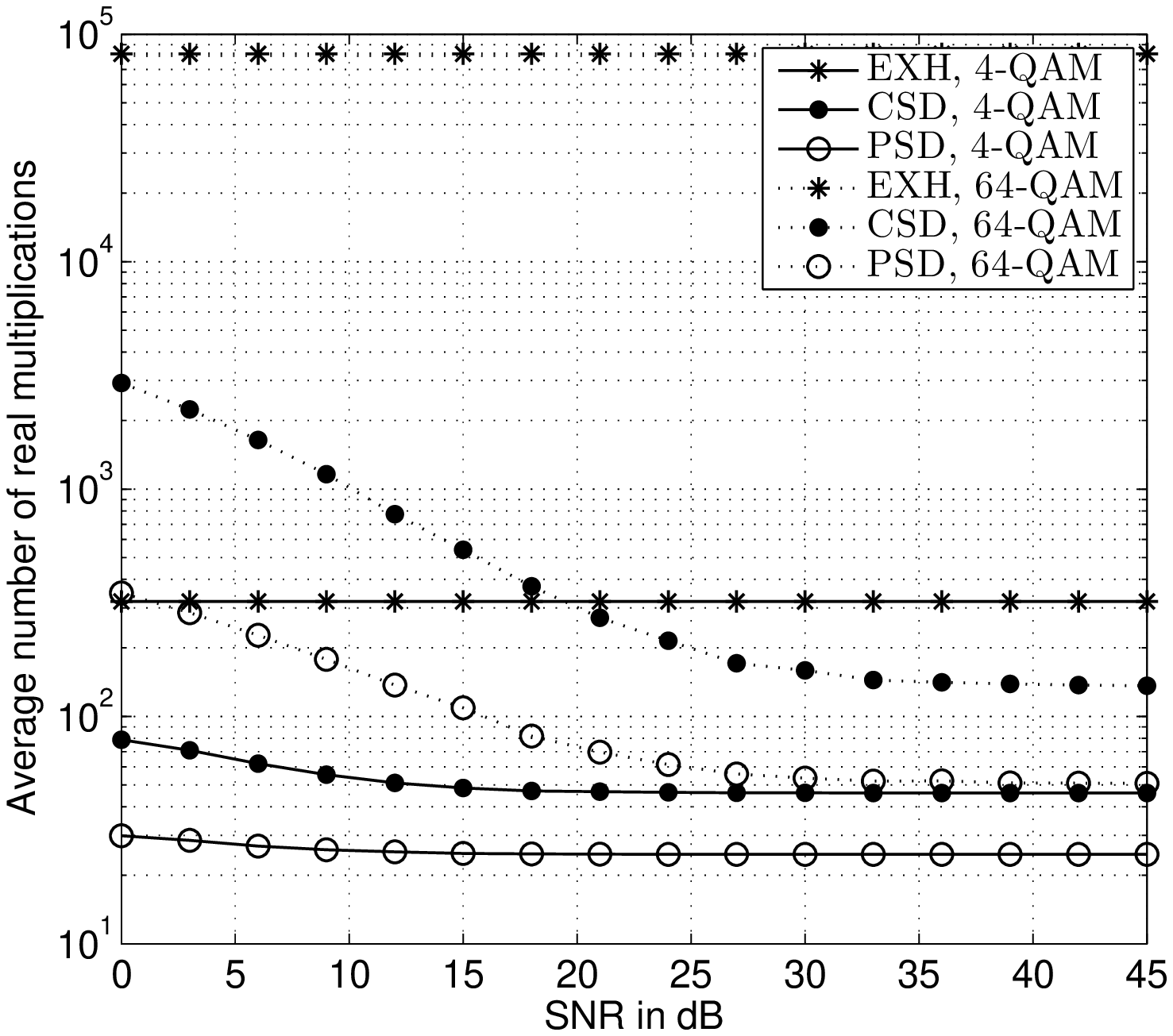} \caption{Average number of real
multiplications vs. SNR for the $2 \times 2$ FPMB systems with
$4$-QAM and $64$-QAM.} \label{fig:mul_2x2_umb}
\end{figure}

\begin{figure}[!t]
\includegraphics[width =
\sizefig\linewidth]{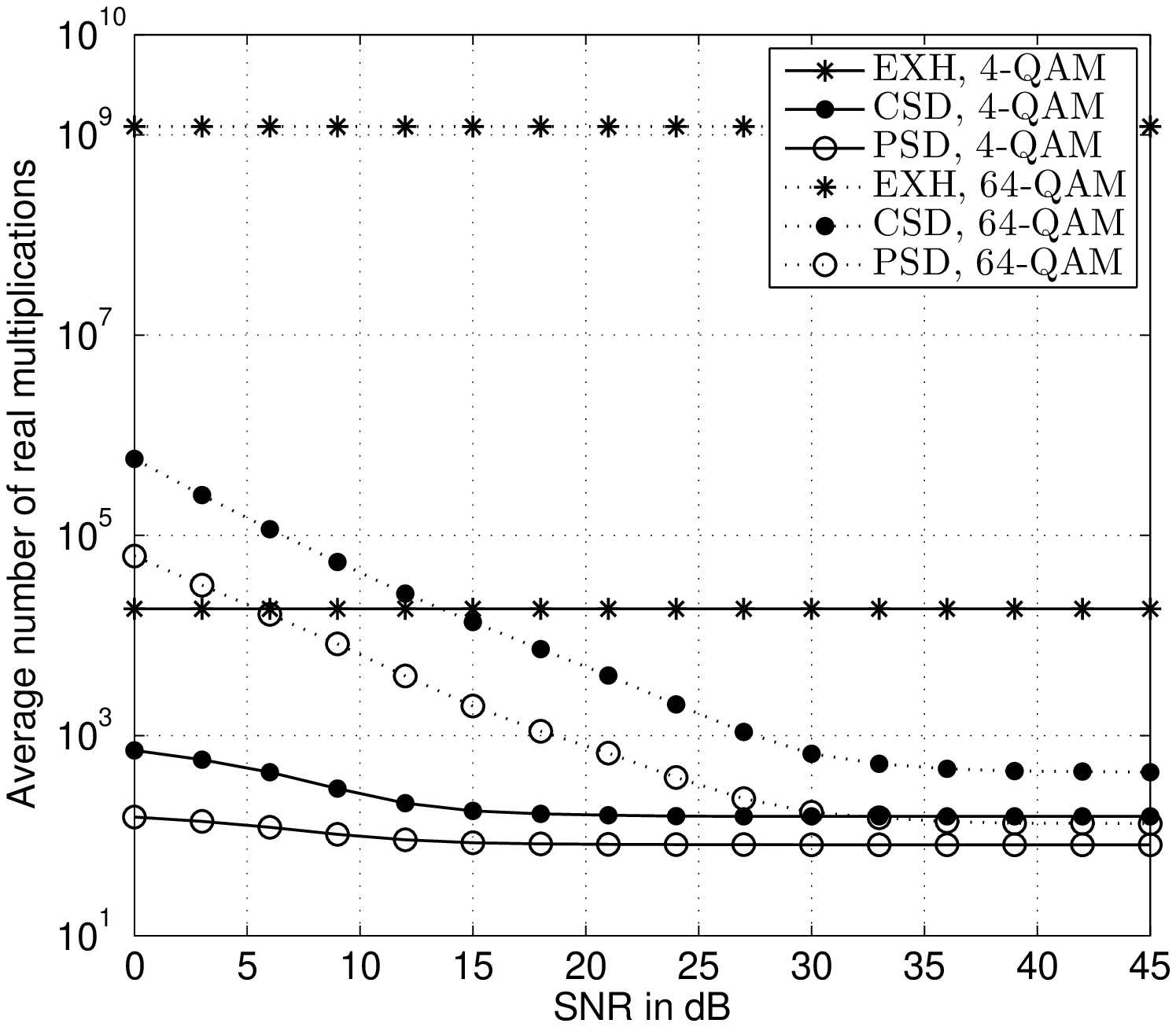} \caption{Average number of real
multiplications vs. SNR for the $4 \times 4$ FPMB systems with
$4$-QAM and $64$-QAM.} \label{fig:mul_4x4_umb}
\end{figure}
\fi

\subsection{BICMB-CP} \label{sec:results_coded}
To verify the diversity analysis in Section
\ref{sec:div_anal_BICMB-CP}, \figurename{
\ref{fig:2x2_BICMBvsBICFPMB}} depicts the simulation results for $2
\times 2$, $3 \times 3$, and $4 \times 4$ BICMB and BICMB-FP with
$64$-state convolutional code punctured from $1/2$-rate mother code
with generator polynomials $(133, 171)$ in octal representation. In
\cite{ParkICC09}, we showed the maximum achievable diversity order
of BICMB with an $R_c$-rate convolutional code is $(N-\lceil S \cdot
R_c \rceil+1)(M-\lceil S \cdot R_c \rceil+1)$. In this example, the
maximum achievable diversity order of the three BICMB systems is
$1$. However, \figurename{ \ref{fig:2x2_BICMBvsBICFPMB}} shows that
BICMB-FP achieves the full diversity order for any code rate.

\figurename{ \ref{fig:3x3_4Q_BICPPMB}} depicts the simulation
results of BICMB-PP given in the example of Section \ref{sec:PPMB}.
The diversity orders of the BICMB systems, $\mathcal{T}_1$ and
$\mathcal{T}_2$ are $4$ and $1$, respectively \cite{ParkICC09}.
Comparing the slopes of BICMB-PP with BICMB, we see that the
simulation results match the analysis in Section \ref{sec:PPMB}.

\ifCLASSOPTIONonecolumn
\begin{figure}[!m]
\centering \includegraphics[width =
0.6\linewidth]{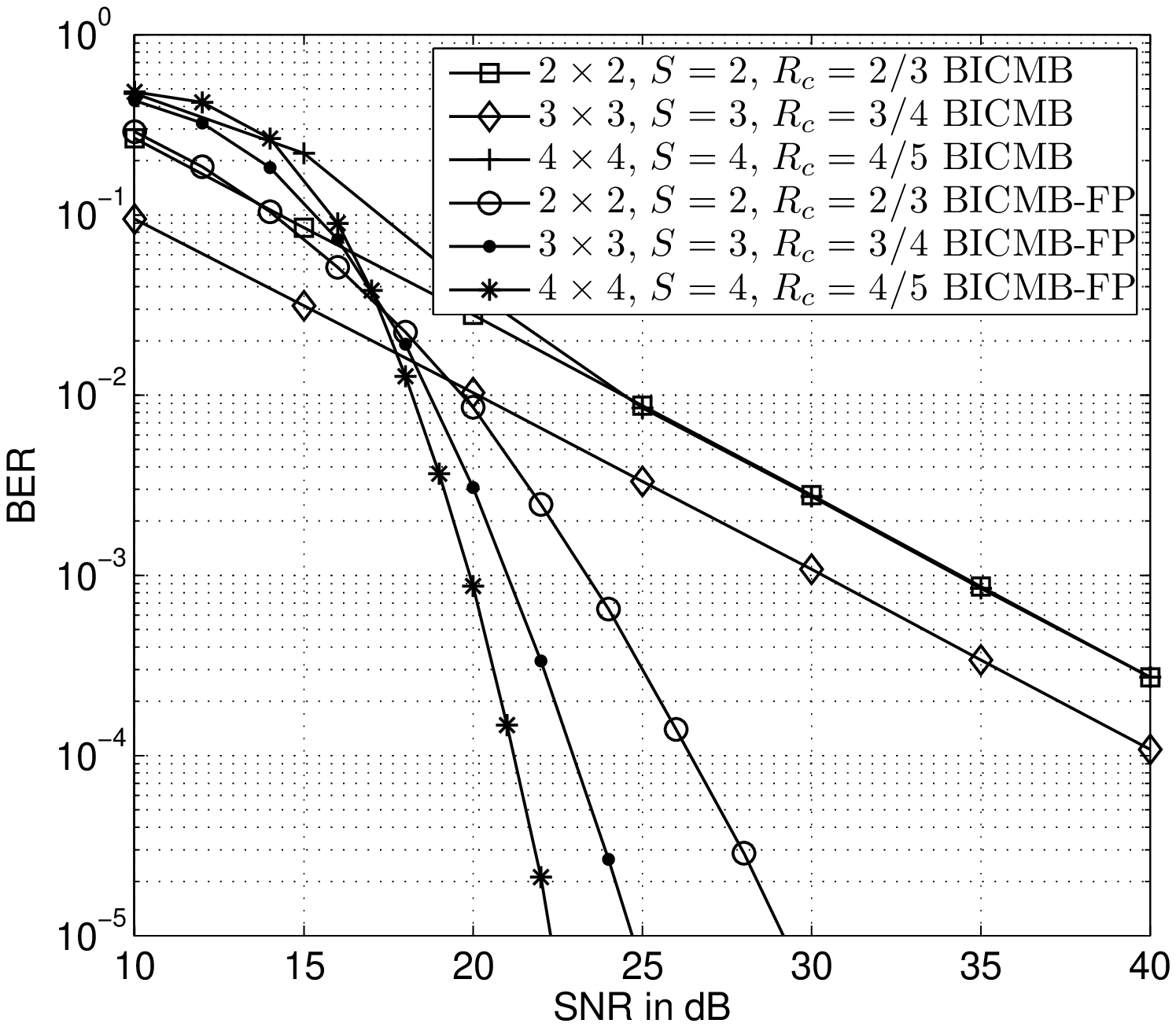} \caption{BER comparison between
BICMB and BICMB-FP with $16$-QAM, and $64$-state punctured
convolutional code.} \label{fig:2x2_BICMBvsBICFPMB}
\end{figure}
\else
\begin{figure}[!t]
\centering \includegraphics[width =
\sizefig\linewidth]{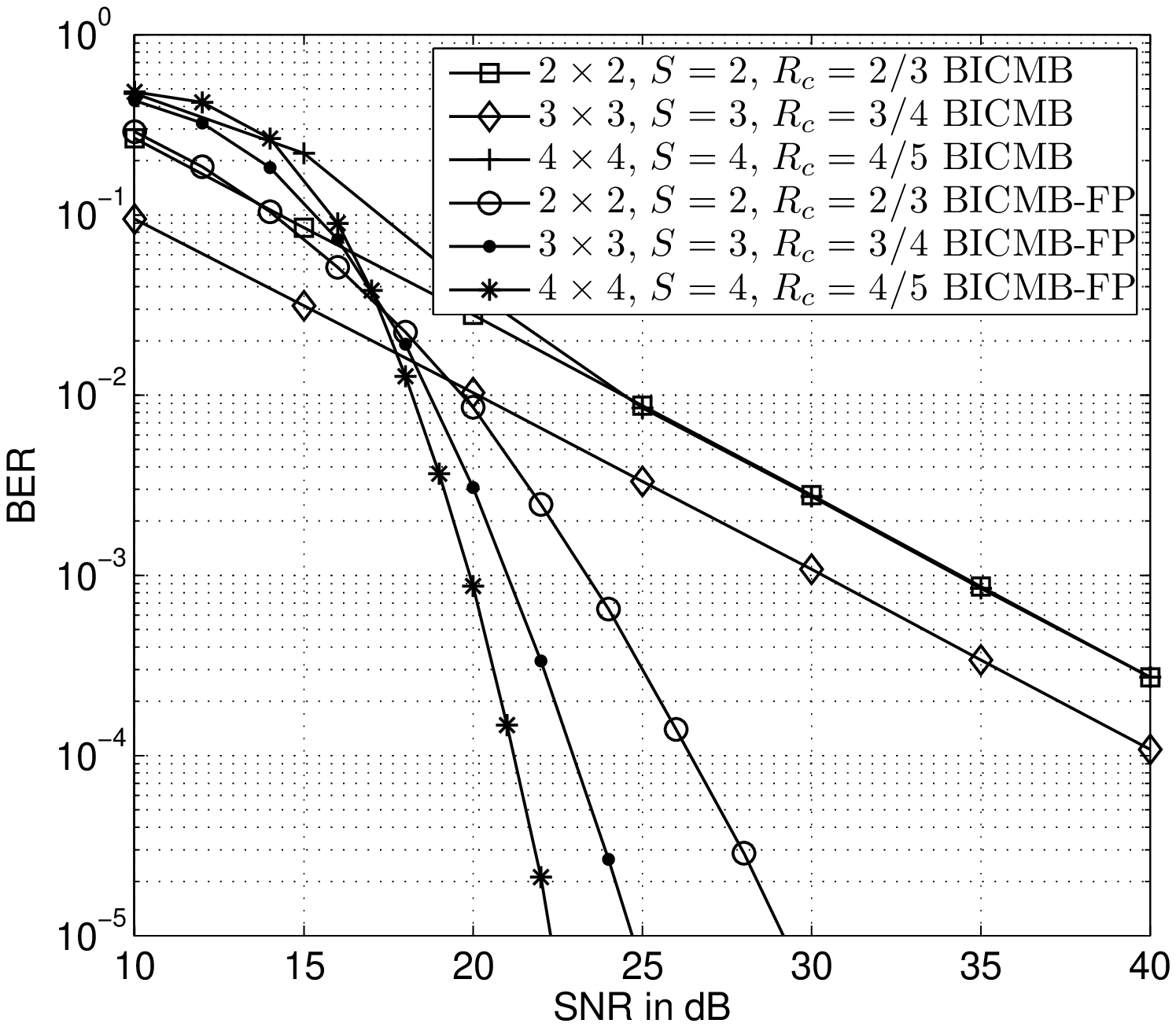} \caption{BER comparison
between BICMB and BICMB-FP with $16$-QAM, and $64$-state punctured
convolutional code.} \label{fig:2x2_BICMBvsBICFPMB}
\end{figure}
\fi

\ifCLASSOPTIONonecolumn
\begin{figure}[!m]
\centering \includegraphics[width =
0.6\linewidth]{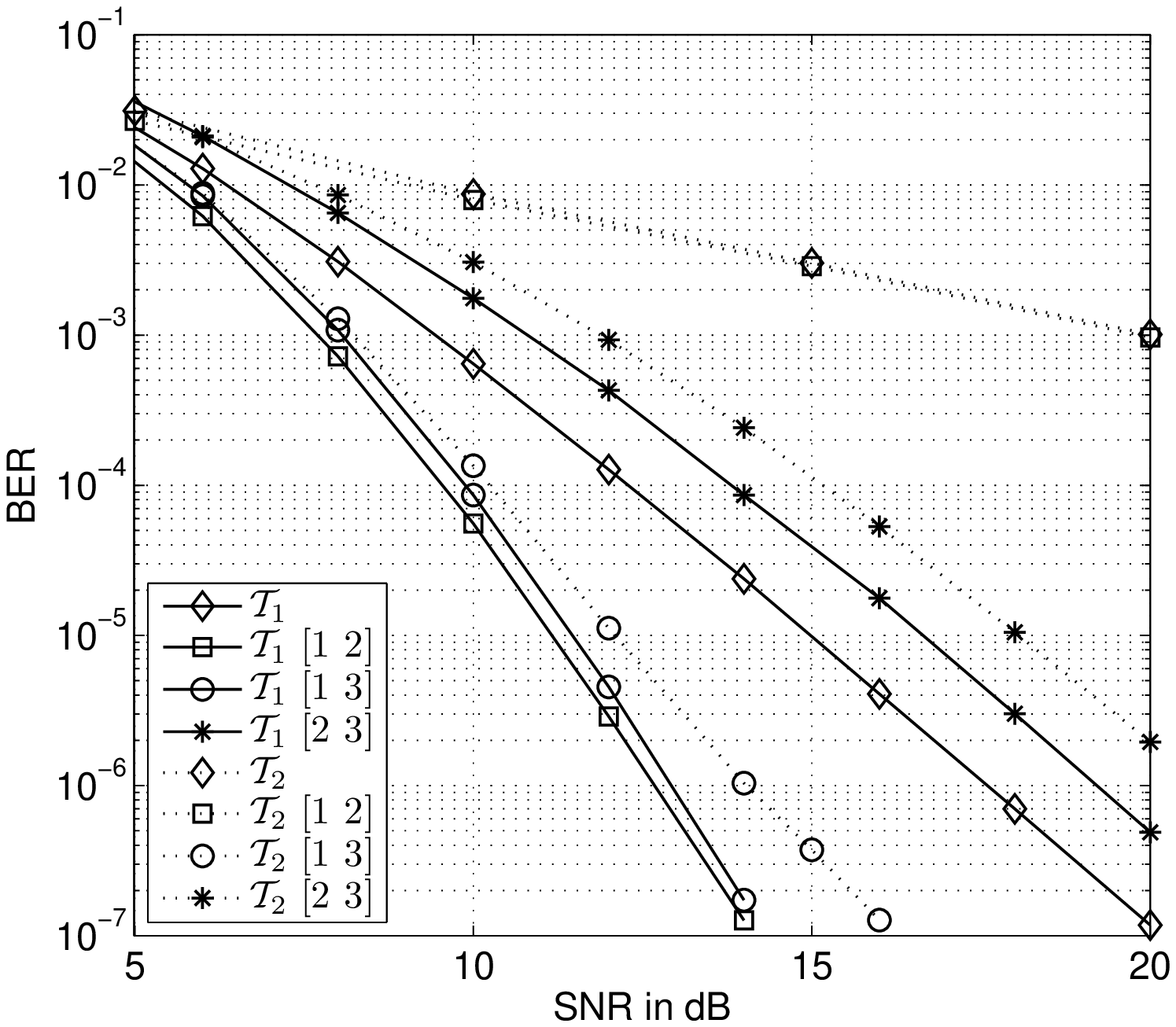} \caption{BER vs. SNR for BICMB-PP
with $3 \times 3$ $S=3$, $4$-QAM, and $4$-state $1/2$-rate
convolutional code.} \label{fig:3x3_4Q_BICPPMB}
\end{figure}
\else
\begin{figure}[!t]
\centering \includegraphics[width =
\sizefig\linewidth]{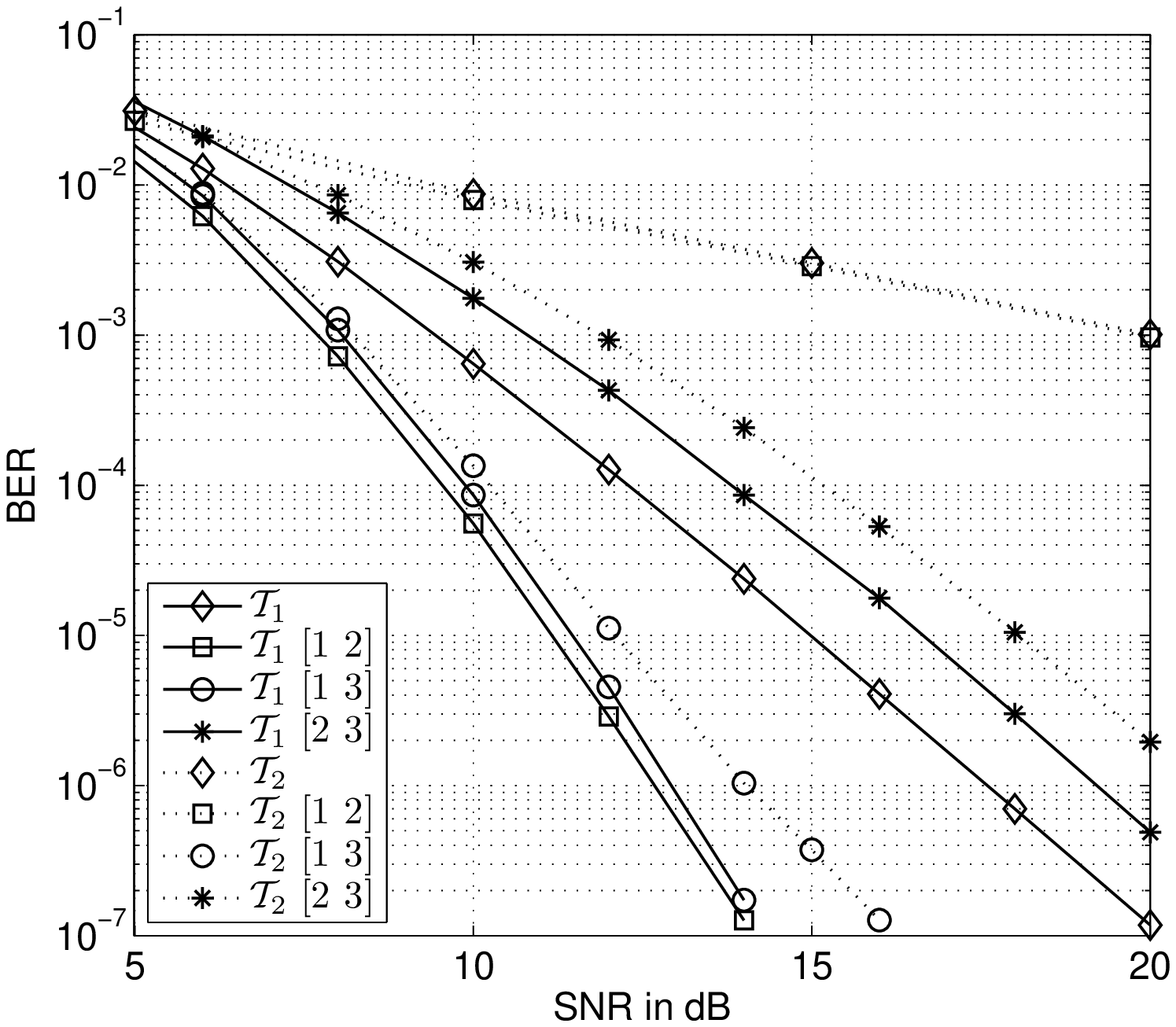} \caption{BER vs. SNR for
BICMB-PP with $3 \times 3$ $S=3$, $4$-QAM, and $4$-state $1/2$-rate
convolutional code.} \label{fig:3x3_4Q_BICPPMB}
\end{figure}
\fi

To verify the proposed sphere decoding technique in this case for
BICMB-FP, we simulated $2 \times 2$ $S = 2$, $64$-state $R_c = 2/3$
BICMB-FP systems, and $4 \times 4$ $S = 4$, $64$-state $R_c = 4/5$
BICMB-FP systems using $4$-QAM and $64$-QAM modulation with Gray
mapping. The average number of real multiplications for acquiring
one bit metric is calculated with receivers employing EXH, CSD, and
PSD. Initial radii for both of CSD and PSD are determined by the
ZF-DFE algorithm. In \figurename{ \ref{fig:mul_2x2_bicmb_cp}}, we
observe that the number of multiplications of CSD for $4$-QAM is
reduced by $0.4$ and $0.5$ orders of magnitude at low and high SNR,
respectively. PSD yields bigger reductions by $1.0$ and $1.1$ orders
of magnitude at low and high SNR, respectively. In the case of
$64$-QAM, reductions between CSD and EXH are $1.5$ and $2.1$ orders
of magnitude at low and high SNR, respectively, while larger
reductions of $2.4$ and $2.9$ are achieved by PSD. \figurename{
\ref{fig:mul_4x4_bicmb_cp}} shows the number of multiplications of
CSD for $4$-QAM decreases by $1.3$ and $1.5$ orders of magnitude at
low and high SNR, respectively. PSD gives bigger reductions by $2.1$
orders of magnitude at low SNR, and $2.3$ at high SNR. For the
$64$-QAM case, reductions between EXH and CSD by $3.2$ and $4.4$
orders of magnitude are observed at low and high SNR, respectively,
while larger reductions by $4.2$ and $5.4$ are achieved by PSD.
Similar to the uncoded case, the complexity reduction becomes larger
as the constellation precoder dimension or the constellation size
becomes larger. One important property of our decoding technique
needs to be emphasized: the substantial complexity reduction
achieved causes no performance degradation.

\ifCLASSOPTIONonecolumn
\begin{figure}[!m]
\centering \includegraphics[width =
0.6\linewidth]{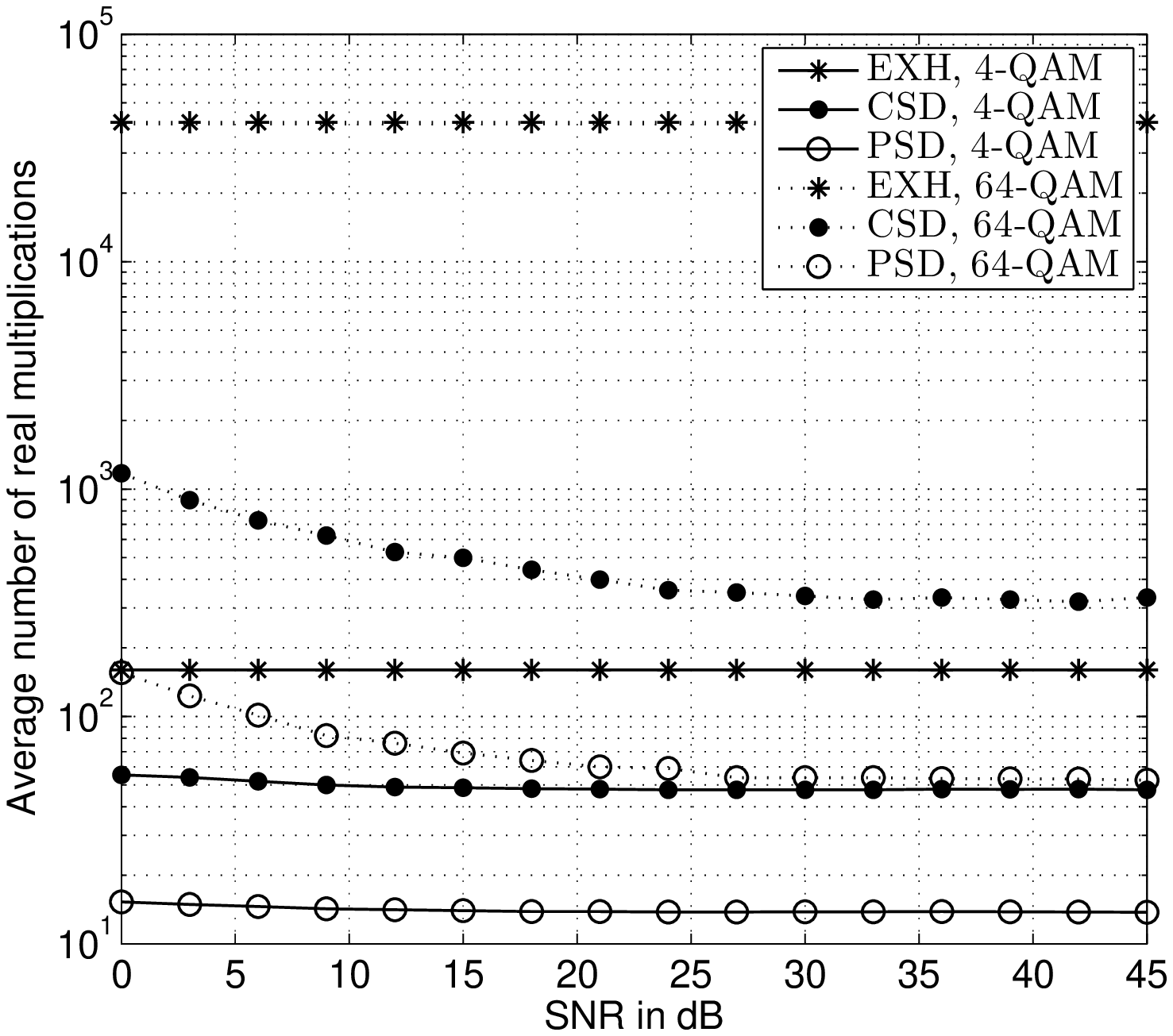} \caption{Average number of real
multiplications vs. SNR for the $2 \times 2$ BICMB-FP systems with
$4$-QAM and $64$-QAM.} \label{fig:mul_2x2_bicmb_cp}
\end{figure}

\begin{figure}[!m]
\centering \includegraphics[width =
0.6\linewidth]{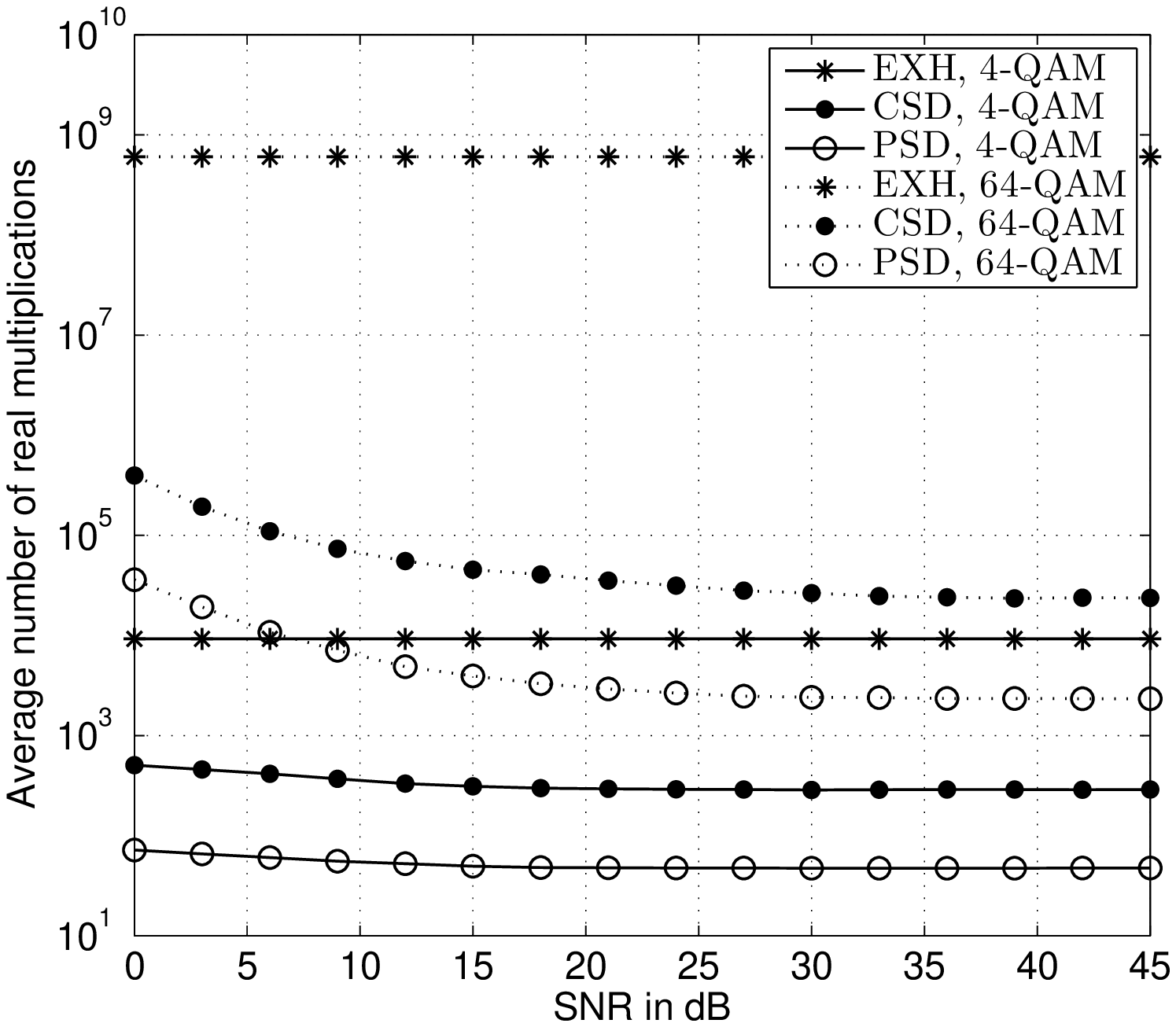} \caption{Average number of real
multiplications vs. SNR for the $4 \times 4$ BICMB-FP systems with
$4$-QAM and $64$-QAM.} \label{fig:mul_4x4_bicmb_cp}
\end{figure}

\else

\begin{figure}[!t]
\includegraphics[width =
\sizefig\linewidth]{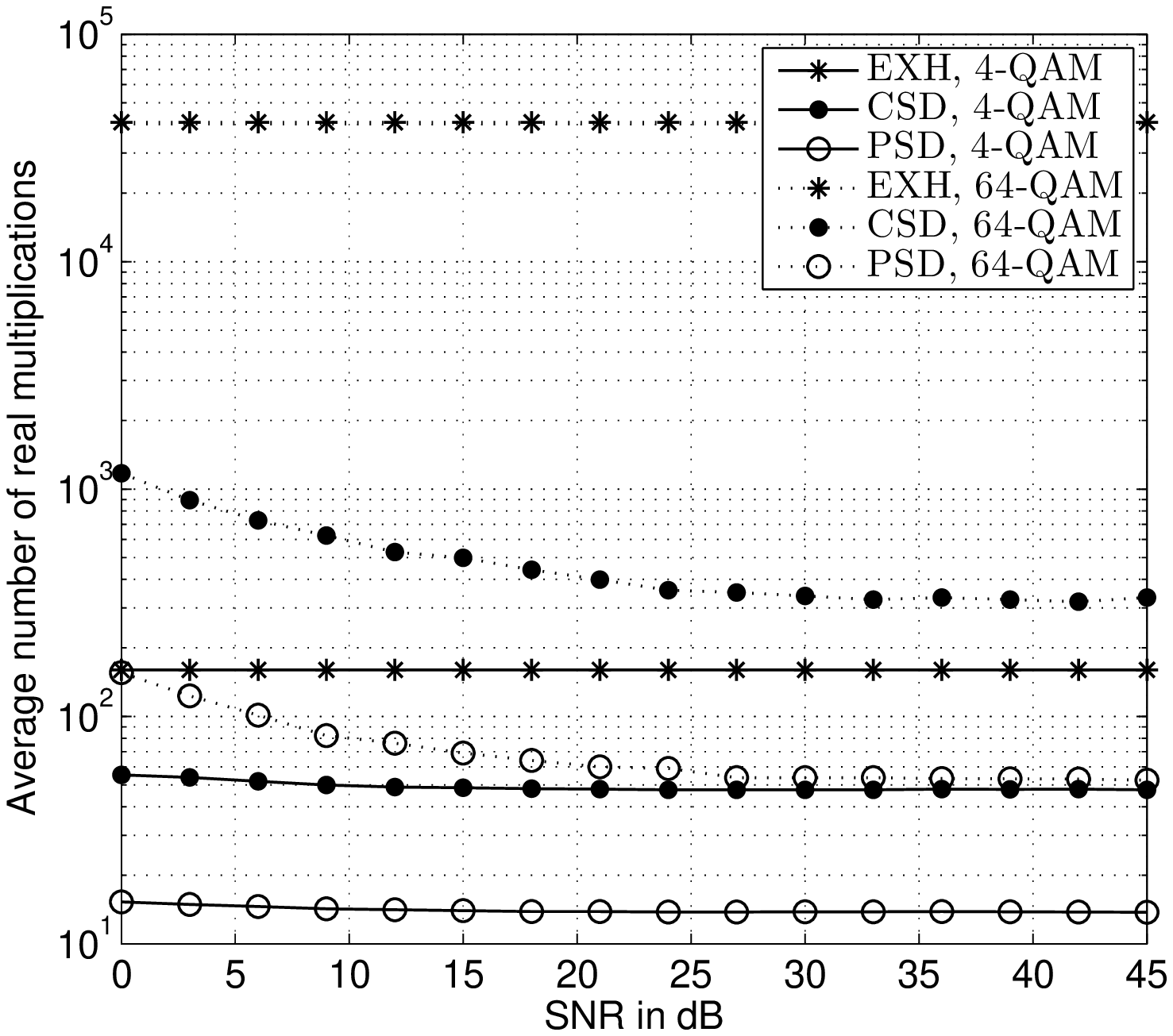} \caption{Average number of
real multiplications vs. SNR for the $2 \times 2$ BICMB-FP systems
with $4$-QAM and $64$-QAM.} \label{fig:mul_2x2_bicmb_cp}
\end{figure}

\begin{figure}[!t]
\includegraphics[width =
\sizefig\linewidth]{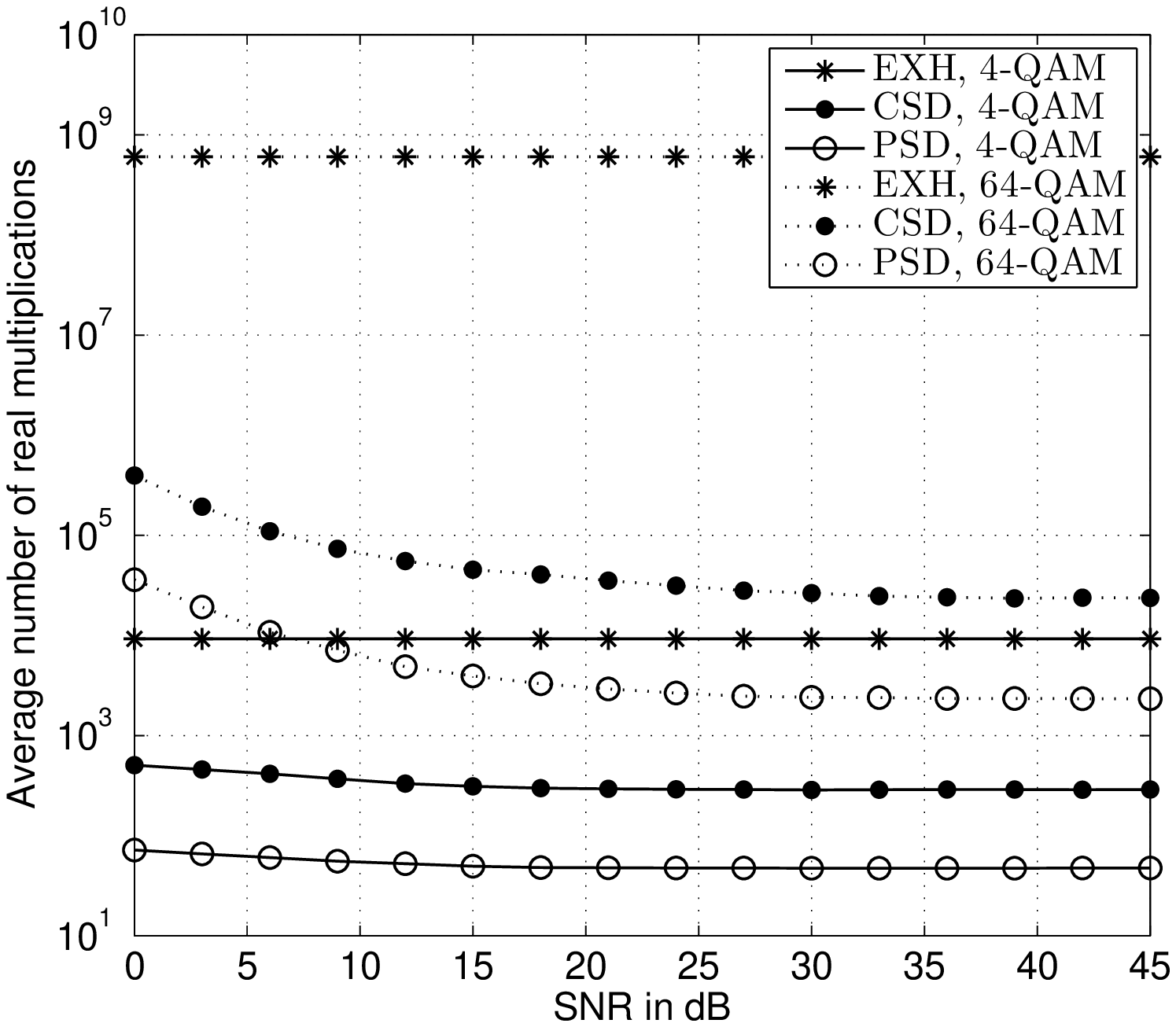} \caption{Average number of
real multiplications vs. SNR for the $4 \times 4$ BICMB-FP systems
with $4$-QAM and $64$-QAM.} \label{fig:mul_4x4_bicmb_cp}
\end{figure}
\fi

\section{Conclusion} \label{sec:conclusion}

In this paper, we proposed constellation precoded multiple
beamforming which achieves the full diversity order in both of the
uncoded and coded MIMO multiple beamforming systems when the channel
information is perfectly available at the transmitter as well as the
receiver, at different levels of spatial multiplexing, including the
maximum $(\min(N,M))$ provided by the $N \times M$ channel.
Diversity analysis was given in both of the multiple beamforming
schemes through the calculation of pairwise error probability. We
provided examples of calculating the diversity orders of various
multiple beamforming systems and simulation results supporting the
analysis. A sphere detection algorithm which improves the complexity
was proposed so that constellation precoded multiple beamforming can
be considered as a practical implementation for MIMO systems
requiring high throughput with the full diversity order. The
proposed SD algorithm in this paper can be applied to any MIMO
system.

\bibliographystyle{IEEEtran}
\bibliography{IEEEabrv,CPMB.bbl}

\begin{thebibliography}{10}
\providecommand{\url}[1]{#1}
\csname url@samestyle\endcsname
\providecommand{\newblock}{\relax}
\providecommand{\bibinfo}[2]{#2}
\providecommand{\BIBentrySTDinterwordspacing}{\spaceskip=0pt\relax}
\providecommand{\BIBentryALTinterwordstretchfactor}{4}
\providecommand{\BIBentryALTinterwordspacing}{\spaceskip=\fontdimen2\font plus
\BIBentryALTinterwordstretchfactor\fontdimen3\font minus
  \fontdimen4\font\relax}
\providecommand{\BIBforeignlanguage}[2]{{%
\expandafter\ifx\csname l@#1\endcsname\relax
\typeout{** WARNING: IEEEtran.bst: No hyphenation pattern has been}%
\typeout{** loaded for the language `#1'. Using the pattern for}%
\typeout{** the default language instead.}%
\else
\language=\csname l@#1\endcsname
\fi
#2}}
\providecommand{\BIBdecl}{\relax}
\BIBdecl

\bibitem{jafarkhaniBook}
H.~Jafarkhani, \emph{Space-Time Coding: Theory and Practice}.\hskip 1em plus
  0.5em minus 0.4em\relax Cambridge University Press, 2005.

\bibitem{SampathJCOM01}
H.~Sampath, P.~Stoica, and A.~Paulraj, ``Generalized linear precoder and
  decoder design for {MIMO} channels using the weighted {MMSE} criterion,''
  \emph{{IEEE} Trans. Commun.}, vol.~49, no.~12, pp. 2198--2206, December 2001.

\bibitem{palomarTSP03}
D.~P. Palomar, J.~M. Cioffi, and M.~A. Lagunas, ``Joint tx-rx beamforming
  design for multicarrier {MIMO} channels: A unified framework for convex
  optimization,'' \emph{{IEEE} Trans. Signal Process.}, vol.~51, no.~9, pp.
  2381--2401, September 2003.

\bibitem{sengulTC06AnalSingleMultpleBeam}
E.~Sengul, E.~Akay, and E.~Ayanoglu, ``Diversity analysis of single and
  multiple beamforming,'' \emph{{IEEE} Trans. Commun.}, vol.~54, no.~6, pp.
  990--993, June 2006.

\bibitem{OrdonezTSP07}
L.~G. Ordonez, D.~P. Palomar, A.~Pages-Zamora, and J.~R. Fonollosa,
  ``High-{SNR} analytical performance of spatial multiplexing {MIMO} systems
  with {CSI},'' \emph{{IEEE} Trans. Signal Process.}, vol.~55, no.~11, pp.
  5447--5463, November 2007.

\bibitem{akayTC06BICMB}
E.~Akay, E.~Sengul, and E.~Ayanoglu, ``Bit interleaved coded multiple
  beamforming,'' \emph{{IEEE} Trans. Commun.}, vol.~55, no.~9, pp. 1802--1811,
  September 2007.

\bibitem{akayTC06BICMB_arxiv}
\BIBentryALTinterwordspacing
E.~Akay, H.~J. Park, and E.~Ayanoglu, ``On bit-interleaved coded multiple
  beamforming,'' 2008, arXiv: 0807.2464. [Online]. Available:
  \url{http://arxiv.org}
\BIBentrySTDinterwordspacing

\bibitem{ParkICC09}
H.~J. Park and E.~Ayanoglu, ``Diversity analysis of bit-interleaved coded
  multiple beamforming,'' in \emph{Proc. {IEEE ICC `09}}, Dresden, Germany,
  June 2009.

\bibitem{GamalJIT03}
H.~E. Gamal and M.~O. Damen, ``Universal space-time coding,'' \emph{{IEEE}
  Trans. Inf. Theory}, vol.~49, no.~5, pp. 1097--1119, May 2003.

\bibitem{XinJWCOM03}
Y.~Xin, Z.~Wang, and G.~B. Giannakis, ``Space-time diversity systems based on
  linear constellation precoding,'' \emph{{IEEE} Trans. Wireless Commun.},
  vol.~2, no.~2, pp. 294--309, March 2003.

\bibitem{LiuJCOM03}
Z.~Liu, Y.~Xin, and G.~B. Giannakis, ``Linear constellation precoding for
  {OFDM} with maximum multipath diversity and coding gains,'' \emph{{IEEE}
  Trans. Commun.}, vol.~51, no.~3, pp. 416--427, March 2003.

\bibitem{ZhangJCOM07}
W.~Zhang, X.-G. Xia, and P.~C. Ching, ``High-rate full-diversity
  space-time-frequency codes for broadband {MIMO} block-fading channels,''
  \emph{{IEEE} Trans. Commun.}, vol.~55, no.~1, pp. 25--34, January 2007.

\bibitem{GressetGlobecom09}
N.~Gresset and M.~Khanfouci, ``Precoded {BICM} design for {MIMO} transmit
  beamforming and associated low-complexity algebraic receivers,'' in
  \emph{Proc. {IEEE Globecom `08}}, New Orleans, LA, November 2008.

\bibitem{ZimmermannWPMC04}
E.~Zimmermann, W.~Rave, and G.~Fettweis, ``On the complexity of sphere
  decoding,'' in \emph{Proc. {Wireless Personal Multimedia Communications
  (WPMC) `04}}, Abano Terme, Italy, September 2004.

\bibitem{JaldenJSP05}
J.~Jald{\'{e}}n and B.~Ottersten, ``On the complexity of sphere decoding in
  digital communications,'' \emph{{IEEE} Trans. Signal Process.}, vol.~53,
  no.~4, pp. 1474--1484, April 2005.

\bibitem{HanGLOBECOM05}
H.~G. Han, S.~K. Oh, S.~J. Lee, and D.~S. Kwon, ``Computational complexities of
  sphere decoding according to initial radius selection schemes and an
  efficient initial radius reduction scheme,'' in \emph{Proc. {IEEE Globecom
  `05}}, St. Louis, MO, November 2005, pp. 2354--2358.

\bibitem{ChengISCC07}
B.~Cheng, W.~Liu, Z.~Yang, and Y.~Li, ``A new method for initial radius
  selection of sphere decoding,'' in \emph{Proc. {IEEE ISCC `07}}, Aveiro,
  Portugal, July 2007, pp. 19--24.

\bibitem{HassibiJSP05}
B.~Hassibi and H.~Vikalo, ``On the sphere-decoding algorithm {I}. {E}xpected
  complexity,'' \emph{{IEEE} Trans. Signal Process.}, vol.~53, no.~8, pp.
  2806--2818, August 2005.

\bibitem{ZhaoJCOM05}
W.~Zhao and G.~B. Giannakis, ``Sphere decoding algorithms with improved radius
  search,'' \emph{{IEEE} Trans. Commun.}, vol.~53, no.~7, pp. 1104--1109, July
  2005.

\bibitem{WongISCAS02}
K.-W. Wong, C.-Y. Tsui, R.~S.-K. Cheng, and W.-H. Mow, ``A {VLSI} architecture
  of a {K-Best} lattice decoding algorithm for {MIMO} channels,'' in
  \emph{Proc. {IEEE ISCAS `02}}, vol.~3, Scottsdale, Arizona, May 2002, pp.
  273--276.

\bibitem{HuynhISWCS08}
T.-A. Huynh, D.-C. Hoang, M.~R. Islam, and J.~Kim, ``Two-level-search sphere
  decoding algorithm for {MIMO} detection,'' in \emph{Proc. {IEEE ISWCS `08}},
  Reykjavik, Iceland, October 2008.

\bibitem{TangICC04}
J.~Tang, A.~H. Tewfik, and K.~K. Parhi, ``Reduced complexity sphere decoding
  and application to interfering {IEEE} 802.15.3a piconets,'' in \emph{Proc.
  {IEEE ICC `04}}, vol.~5, Paris, France, June 2004.

\bibitem{AzzamGLOBECOM07}
L.~Azzam and E.~Ayanoglu, ``Reduced complexity sphere decoding for square {QAM}
  via a new lattice representation,'' in \emph{Proc. {IEEE Globecom `07}},
  Washington, D.C., November 2007.

\bibitem{ZanellaICC08}
A.~Zanella, M.~Chiani, and M.~Z. Win, ``A general framework for the
  distribution of the eigenvalues of {W}ishart matrices,'' in \emph{Proc. {IEEE
  ICC `08}}, May 2008, pp. 1271--1276.

\bibitem{ParkGlobecom09}
H.~J. Park and E.~Ayanoglu, ``Constellation precoded beamforming,'' in
  \emph{Proc. {IEEE Globecom `09}}, Honolulu, HI, November 2009.

\bibitem{park-2009_arxiv}
\BIBentryALTinterwordspacing
------, ``Constellation precoded beamforming,'' 2009, arXiv:0903.4738v1.
  [Online]. Available: \url{http://arxiv.org}
\BIBentrySTDinterwordspacing

\bibitem{HassibiICASSP02}
B.~Hassibi and H.~Vikalo, ``On the expected complexity of integer least-squares
  problems,'' in \emph{Proc. {IEEE ICASSP `02}}, vol.~2, Orlando, FL, May 2002.

\end{thebibliography}

\end{document}